\documentclass[%
reprint,
superscriptaddress,
frontmatterverbose, 
amsmath,amssymb,aps,
]{revtex4-2}
\usepackage{xcolor}
\usepackage{graphicx}
\usepackage{dcolumn}
\usepackage{bm}
\usepackage{blindtext}
\usepackage{lipsum}
\usepackage{hyperref}
\usepackage[mathlines]{lineno}
\usepackage{amsmath}
\usepackage{multirow}
\usepackage{graphicx}
\usepackage{tabularx}
\usepackage{setspace}
\usepackage{makecell}
\usepackage{natbib}

\usepackage[
scale=0.85, marginratio={1:1, 2:3}, ignore all,
margin=0.65in,
]{geometry}

\begin{document}

\preprint{APS/123-QED}

\title{Parts-per-billion Trace Element Detection in Anhydrous Minerals by Micro-scale Quantitative NMR}

  \author{Yunhua Fu}
  \affiliation{School of Earth and Space Sciences, Peking University, Beijing, China}
  \affiliation{Center for High-Pressure Science and Technology Advance Research, Beijing, China}
 
 \author{Renbiao Tao}
  \email{renbiao.tao@hpstar.ac.cn}
   \affiliation{Center for High-Pressure Science and Technology Advance Research, Beijing, China}  

      \author{Lifei Zhang}
  \affiliation{School of Earth and Space Sciences, Peking University, Beijing, China} 
   
    \author{Shijie Li}
  \affiliation{Center for Lunar and Planetary Sciences, Institute of Geochemistry, Chinese Academy of Sciences, Guiyang, China} 
 
   \author{Yanan Yang}
  \affiliation{State Key Laboratory of Isotope Geochemistry, Guangzhou Institute of Geochemistry, Chinese Academy of Sciences, Guangzhou, China}

  \author{Dehan Shen}
  \affiliation{Center for Lunar and Planetary Sciences, Institute of Geochemistry, Chinese Academy of Sciences, Guiyang, China} 

\author{Zilong Wang}  
  \affiliation{School of Earth and Space Sciences, Peking University, Beijing, China}

  \author{Thomas Meier}
  \email{thomas.meier@sharps.ac.cn}
  \affiliation{Shanghai Key Laboratory MFree, Institute for Shanghai Advanced Research in Physical Sciences, Pudong, Shanghai, 201203}
  \affiliation{Center for High-Pressure Science and Technology Advance Research, Beijing, China}

\date{\today}

\begin{abstract}
Nominally anhydrous minerals (NAMs) composing Earth's and planetary rocks incorporate microscopic amounts of volatiles. However, volatile distribution in NAMs and their effect on physical properties of rocks remain controversial. Thus, constraining trace volatile concentrations in NAMs is tantamount to our understanding of the evolution of rocky planets and planetesimals. Here, we present a novel approach of trace-element quantification using micro-scale Nuclear Magnetic Resonance (NMR) spectroscopy. This approach employs the principle of enhanced mass-sensitivity in NMR microcoils formerly used in \textit{in-situ} high pressure experiments. We were able to demonstrate that this method is in excellent agreement with standard methods across their respective detection capabilities. We show that by simultaneous detection of internal reference nuclei, the quantification sensitivity can be substantially increased, leading to quantifiable trace volatile element amounts of about $50$ wt-ppb measured in a micro-meter sized single anorthitic mineral grain, greatly enhancing detection capabilities of volatiles in geologically important systems.  
 \end{abstract}
\maketitle

\section{Introduction}

Hydrogen and halogens are important tracers in the geodynamic process of Earth and other planets owing to their high volatility and incompatibility\cite{Yoshino2018}. Hydrogen is predominantly stored in hydrous minerals but also present in NAMs in amounts ranging from $\sim10^{-1}$ to $10^{5}$ wt-ppm \cite{Ohtani2015} coordinated as hydroxyl groups ($OH^{-}$), water molecules ($H_2O$) or occasionally methane ($CH_4$) and molecular hydrogen($H_2$) in grain boundaries, crystalline vacancies or in grain boundary interstitials \cite{Demouchy2016}. Parallel to hydrogen, significant amounts of halogens (e.g. fluorine) and phosphorus can be absorbed in minerals due to their often comparable ionic charge and radius. Previous studies have shown that terrestrial pyroxenes can incorporate up to $50$ wt-ppm of fluorine, while clinopyroxenes from 4 Vesta can contain as little as $60$ wt-ppb\cite{Sarafian2019, Beyer2012}. Phosphorus serves as the major element in apatite as well as is present in NAMs as the trace element, whose abundance could differ by 4-5 orders of magnitude\cite{Macia2005, Pasek2019, Kumamoto2017, Walton2021}. Albeit being abundant only in trace amounts, the presence of hydrogen and fluorine is known to significantly alter physical properties of mantle minerals\cite{Schmandt2014} with large-scale impacts for lithospheric mantle stability \cite{Peslier2010, Xia2019, Beyer2012}.
\\
\begin{figure}[htb]
\includegraphics[width=0.8\columnwidth]{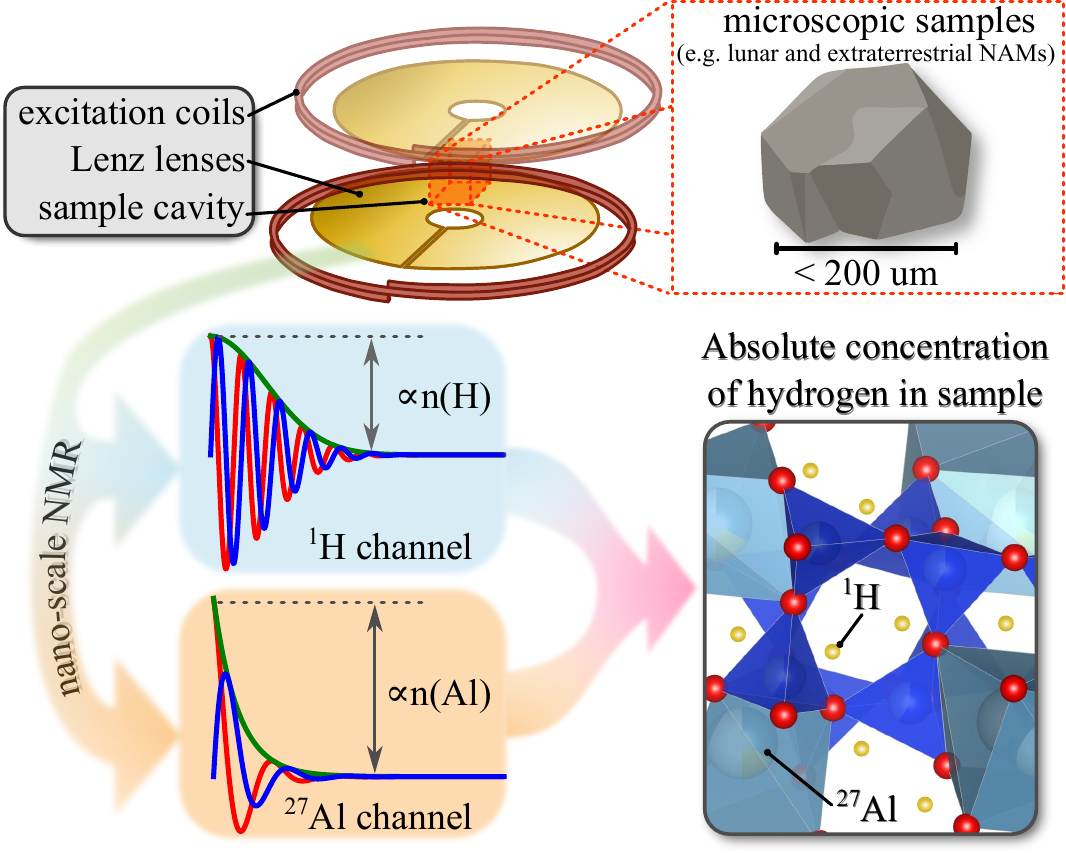}
\caption{\textbf{Schematic representation of the quantification method using NMR spectroscopy.} Due to small sample sizes, detection of the NMR transient after spin excitation is accomplished by the use of magnetic flux-focusing Lenz lenses. Since signal intensities are directly proportional to the number of spins, this method provides a direct measurement of the spin density (spins per sample volume) in a microscopic specimen. Moreover, abundant NMR-active nuclei in the same sample (e.g. $^{27}Al$) can be employed as internal reference standards, making this method fully stand-alone and non-destructive for trace-element quantification.}
\label{fig:principle}
\end{figure}
\begin{table*}[ht!]
\centering
\caption{Summary of $H$ or $H_2O$ concentrations from $\mu$Q-NMR and other methods of all sample systems}
\resizebox{0.7\textwidth} {!}{
    \begin{tabular} {cccccc}
         \hline
\multirow{2}{*}{} & \multirow{2}{*}{sample}  & \multicolumn{2}{c}{\makecell{$n$ from \\known sources}}&\multicolumn{2}{c}{\makecell{$n$ from\\ $\mu$Q-NMR}} 
\\
    &   &  method  &  (wt-ppm)  & ref. & (wt-ppm) 
\\
           \hline
\multirow{6}*{\makecell{pyolitic and\\ rhyolitic glasses}} & OA-1 & NanoSIMS & 1214(30) &- & 1321(66)\\
 & RH-4 & NanoSIMS &12742(250) & -&10367(25)
\\
 & RH-9 &FTIR& 63500(600)&\cite{Tu2023} &58098(30)
 \\
 & PYR 321-01 & NanoSIMS &700(100)&-&642(50)
 \\
 & PYR 321-02& NanoSIMS &900(200)&-&783(25)
 \\
 & PYR 321-03 & NanoSIMS&700(100)&-&570(60)
 \\
 \hline
 \multirow{3}*{NAMs}& stishovite & NanoSIMS &6200(1500) &\cite{Li2023}&4300(100)
\\
 & bridgmanite &NanoSIMS&1100(150)&-&1163(120)
 \\
 \hline
  \multirow{9}*{\makecell{$LaH_3/Al$ mixtures}}& $$H/Al$$-5 &\multirow{9}*{Mass balance}&18957(190)&\multirow{9}*{-}&18829(40)
\\
& $H/Al$-2.5 &&17185(320)&&17208(60)
 \\
& $H/Al$-1  &&13428(490) &&13780(220) 
\\
& $H/Al$-0.5  &&9847(520) && 9173(590) 
\\
& $H/Al$-0.25 &&6426(440) &&6457(890)
\\
& $H/Al$-0.1 &&3147(260) &&3055(540)
\\
& $H/Al$-0.05 &&1702(150) &&1544(150) 
\\
& $H/Al$-0.025 &&887(80) &&784(130) 
\\
& $H/Al$-0.01 &&364(30) &&361(30)
\\
 \hline
 \multirow{7}*{\makecell{Metal hydrides \\ in DACs }}& $FeH$ &\multirow{7}*{XRD + calc.}& 17525(4300)&\multirow{7}*{\cite{Meier2023}}&17187(3440)
\\
 & $CuH_{0.15}$ && 2356(580) &&2042(470)
\\
 & $Cu_2H$ && 7809(1930) &&4077(1170)
\\
 & $CuH$ && 15489(3810) &&17179(910)
\\
& $H_5S_6$ && 35957(8680) &&42271(5710)
\\
 & $YH_2$ && 21972(5370) &&26685(5220)
\\
 & $YH_3$ && 32580(7890) &&35225(6490)
\\
\hline
\end{tabular}
}
    \label{tab I}       
  \end{table*}
Almost all NAMs constituting Earth's crust and mantle are known to incorporate variable and measurable amounts of volatiles \cite{Keppler2006}, adding up to significant reservoirs in the deep Earth \cite{Peslier2017}. While experimental studies have provided insights into stored global amounts of those volatiles in the upper mantle and transition zone, estimates of hydrogen and fluorine reserves in lower mantle regions remain controversial\cite{Fei2017, Mao2012, Huang2005, Yoshino2018}. It is widely believed that lower mantle minerals are capable of storing up to a $4.5$-fold surface water within their crystal structures \cite{Keppler2006a, Gruninger2017, FelixV.Kaminsky2018}. However, trace volatile element quantification methods are often prone to systematic errors and depend on cross-calibration with other methods\cite{Johnson2003, Tu2023}. 
\\
Due to challenges constraining absolute trace element concentrations, origin and accretion mechanisms of volatiles on Earth and other planetary bodies remain controversially discussed \cite{Newcombe2023a, Beyer2012, Sarafian2019}.
Contemporary consensus within the geoscience community holds that either hydrogen incorporation in primitive minerals - formed during nebular in-gassing and early planetary accretion -  or extra-terrestrial bombardment of carbonaceous chondrites and cometary materials, originating from outer solar system during later stages of planet formation, are dominant hydrogen influx mechanisms\cite{Broadley2022, Newcombe2023a}. Recent studies suggest that differentiated parent bodies have been invoked as significant volatile sources for Earth, as indicated by comparisons of water concentrations among Earth and other precursors\cite{ Piani2020, Peterson2023, Peterson2023b}. Thus, using comparative planetology to understand volatile delivery in the solar system and the Earth's deep water cycle is fundamental. It is imperative to quantify hydrogen and other volatile trace element concentrations in terrestrial and extra-terrestrial minerals using high-mass-sensitivity methods.
\\
Techniques for determining volatile contents have recently been developed with mass sensitivities allowing for trace volatile element detection in the wt-ppm range. Fourier Transform Infrared Spectroscopy (FTIR) is a commonly used method in geo-material research\cite{Hauri2002} capable of detecting reliable low hydrogen abundance relying on accurate identification of optical orientation\cite{Hu2015, Koga2003}. However, being primarily sensitive to molecular vibrations, FTIR spectroscopy detects molecular sub-groups (i.e. hydroxyl-groups or molecular water) as opposed to individual elements or interstitial trace elements.
Moreover, molecular \textit{H$_2$} is particularly difficult to detect using FTIR due to its nonpolarity\cite{Moine2020}, rendering a reliable quantification dependent on numerous experimental, and often varying, parameters (e.g.molar absorption coefficient)\cite{ Johnson2003, Mosenfelder2015}.
\\
Secondary Ion Mass Spectrometry (SIMS) on the other hand has become a routine choice for trace volatile element quantification in natural or synthetic minerals, with mass-sensitivities comparable to that of FTIR techniques\cite{Hu2015, Hui2017}. However, the requirement for standard samples and matrix effects make SIMS greatly dependent on sample quality and thus its volatile detection limits are also linked to numerous technical parameters and prone to systematic errors\cite{Aubaud2007, KATAYAMA2006, Johnson2003, Mosenfelder2015}.  
\\
Nuclear Magnetic Resonance spectroscopy (NMR) is widely considered as one of the most widespread spectroscopic methods in contemporary natural science laboratories, with a plethora of applications spanning physics, chemistry, medicine and materials science; alas its use in geo-science has historically been limited. Despite being predominantly recognized for its versatile qualitative feature, e.g. in atomic and electronic structure determination\cite{Seifrid2020}, the highly quantitative nature of NMR is often used in metabolomics\cite{Crook2020}  or quantitative chemistry\cite{Khalil2021}.
\\
Quantitative NMR relies on the principle of reciprocity \cite{Hoult1976}, tying signal intensities to the spin density in a sample (e.g. $^1H$, $^{27}Al$,$^{19}F$ and $^{31}P$ per unit volume), see Figure \ref{fig:principle}. Modern efforts were largely sparked by substantial advances in detection sensitivities for several analytical techniques\cite{McCubbin2015}. Detection limits of NMR are typically limited to a few weight percent\cite{Li2020}, due to the weak induced nuclear polarisation of the nuclear spin system. However, it has been argued that mass-sensitivity can be enhanced significantly using radio-frequency (RF) micro-coils\cite{Peck1995} compared to commonly used macroscopic NMR solenoids\cite{Olson1995, Schlotterbeck2002, Schroeder2006}.\\
\begin{figure}[ht]
\centering    
\includegraphics[width=1\columnwidth]{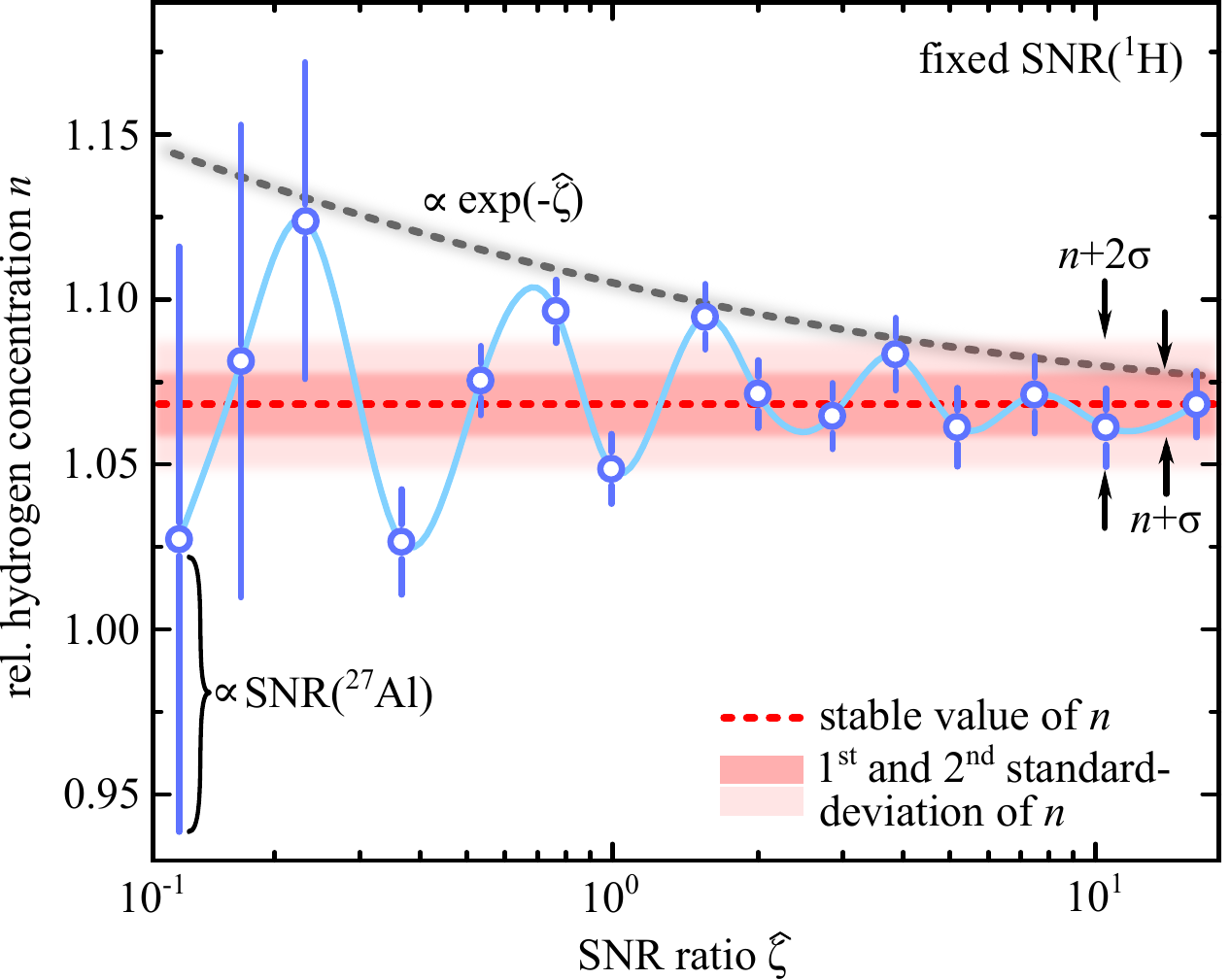}
\caption{\textbf{Calibration of hydrogen concentration using Nuclear Magnetic Resonance Spectroscopy.}  Relative hydrogen concentration \textit{n} as a function of the SNR ratio $\hat{\zeta}$ at fixed SNR($^1$H). Usually, reliable values of \textit{n}, within $\pm 1\sigma$ confidence levels, are approached after $\hat{\zeta}\approx 10^1$ has been achieved.
}
\label{fig: H calibration}
\end{figure}
\\
Recent advances in \textit{in-situ} high pressure NMR in diamond anvil cells using magnetic flux tailoring techniques enhanced detection limits, and thus mass-sensitivities, by several orders of magnitude\cite{Meier2018b, Meier2019b} compared to commercially available NMR systems. With these novel detection methods, it could be shown that the faint nuclear induction signal stemming from samples as small as $\approx 10~pl$ can be detected with high fidelity\cite{Meier2019}.    
\\
Amongst most volatile trace elements significant for geo-scientific research, $^1H$,$^{19}F$ and $^{31}P$ are so-called "high $\gamma_n$ nuclei" often yielding strong signal strength and generally simple line-shapes due to their single quantum nuclear spin being devoid of quadrupolar interactions. Therefore spin system made from those nuclei can be detected even at low concentrations. 
\\
Here we introduce a refined trace element quantiﬁcation method using Micro-scale NMR. Using well known and characterized samples, we show that this new method results in hydrogen concentrations in very good agreement with FTIR and Nano-SIMS techniques up to their respective detection limits. Furthermore, we show that by simultaneous detection of well-refined reference spin systems present in the same sample (e.g. $^{27}$Al)(Figure \ref{fig:principle}), the volatile trace element detection limit can be as low as $40$ wt-ppb for a wide array of geo-physically important samples and elemental probes. Due to its wide applicability and orders of magnitude better detection limits, $\mu$Q-NMR has the potential to revolutionize contemporary quantitative geo-science.  
\section{Results}

\begin{figure}[h!]
\centering    
\includegraphics[width=.9\columnwidth]{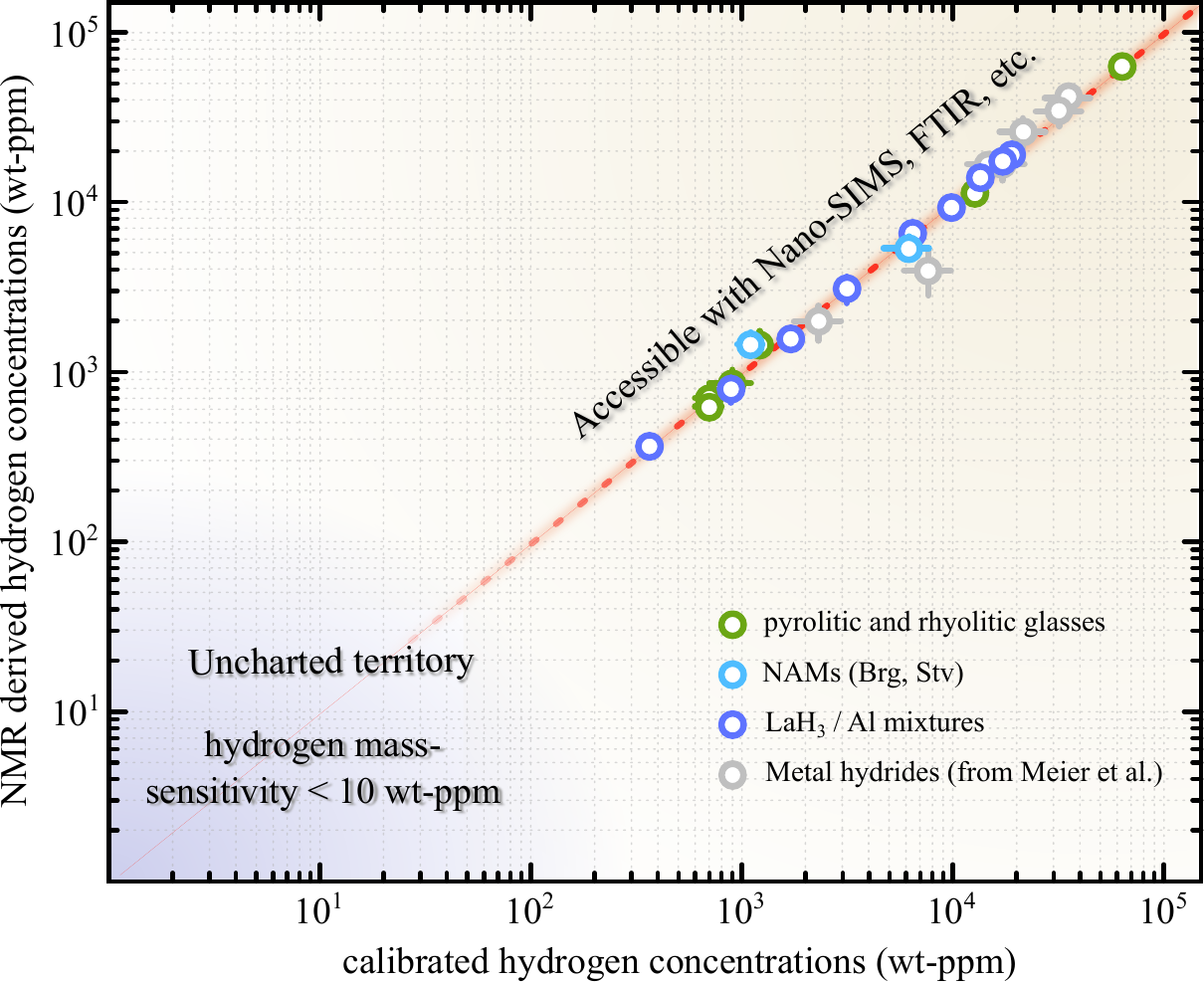}
\caption{\textbf{Benchmarking of \textit{$n(^1H)$} obtained from $\mu$Q-NMR.}  The investigated hydrogen or water concentrations range over three orders of magnitude from $\sim 10^2$ to $10^5$ wt-ppm. NMR derived values of \textit{$n(^1H)$} have been compared to concentrations derived from standard methods. The red line is a guide to the eye for a 1:1 correlation.
}
\label{fig: H comparison}
\end{figure}

\textbf{$\mu$Q-NMR method:}\\
Figure \ref{fig:principle} illustrates schematically the method of our employed quantitative NMR approach. Microscopic samples are loaded into the (roughly $300~\mu m\times300~\mu m\times300~\mu m$) sample cavity located in the center of a double Lenz lens micro-resonator \cite{Jouda2017}. Driving coils were either designed from simple solenoids for low frequency applications or using PCB-based single-turn NMR coils terminated and impedance matched to resonate close to $400$ MHz\cite{Meier2018, Meier2022}.
\\
Without loss of generality, the here presented self-consistent reference-sample-free method of trace element quantification relies on the presence of at least two NMR-active spin sub-systems, i.e. \textit{A} and \textit{B}. With sub-system \textit{A} being the to-be-quantified trace element and subsystem \textit{B} being a more abundant subsystem. One can define the ratio \textit{$\hat{\zeta}$} of the signal-to-noise ratios of each individual nuclear spin sub-system \textit{$\zeta_A$} and \textit{$\zeta_B$}. As both subsystems are part of the same sample, detected within the same resonator, \textit{$\hat{\zeta}$} becomes independent of most experimental parameters, like sample-to-coil filling factors or coil dimensions. Thus one easily obtains the spin density ratio \textit{$\hat{n}$=n(A)/n(B)} by comparison of  \textit{$\zeta_A$} and \textit{$\zeta_B$} after correcting for nuclear spin quantum numbers \textit{I}, gyromagnetic ratios $\gamma_n$  and digitisation frequencies ($\Delta f=1/DW$):

\begin{equation}
    \hat{n}=\frac{\zeta_A}{\zeta_B}\cdot\left(\frac{\gamma_A}{\gamma_B}\right)^{\frac{11}{4}}\cdot\Theta(I)\cdot\Delta_{DW}
    \label{1}
\end{equation}
where $\zeta_{A,B}$ denote the respective time domain SNR for each spin-subsystem and $\gamma_{A,B}$ are the respective gyromagnetic ratios. 
The function $\Theta(I)$ corrects for non-equivalent nuclear spin quantum numbers with $\Theta(I)=(I_{A}(I_{A}+1))/(I_{B}(I_{B}+1))$ and $\Delta_{DW}$ corrects for different digitisation frequencies (i.e. dwell times $DW$), $\Delta_{DW}=\sqrt{DW_B/DW_A}$).
\\
The obtained concentration ratio (eq. \ref{1}) has been shown to be workable in \textit{in-situ} high-pressure NMR experiments in diamond anvil cells for metal hydrides \cite{Meier2023} where simultaneous detection of hydrogen and an NMR-active parent metal nuclear spin subsystem were possible. A systematic \textit{a-priori} quantification attempt (following eq. \ref{1}) of hydrogen or other volatiles in geophysically important materials has not been conducted so far.
\\
Considering that alumino-silicate minerals, which are abundant in the crust and mantle, exhibit large amounts of \textit{Al}, the $^{27}Al$ nuclear spin sub-system is a natural choice to be used as an internal reference for volatile trace element quantification for a plethora of geo-samples. It should be noted at this point that, in order for eq. \ref{1} to yield actual concentration numbers of trace elements, the total amount of alumina in each sample needs to be determined, i.e. via electron microprobe analysis.
\\
Throughout this study, a customized python script for calculation of the quantification was used, which is available to the interested reader upon request. 
\\
\textbf{Benchmark and calibration:}\\
In order to thoroughly benchmark $\mu$Q-NMR based trace element detection, we primarily focused on spin pairs of hydrogen as the to-be-quantified atom species and aluminum as internal references. First benchmark tests were conducted on powder mixtures of Lanthanum trihydride and metallic Aluminium, with the overall goal of covering two orders of magnitude of Hydrogen-to-Aluminium abundance ratios, $H/Al$. Nine samples, from $H/Al$ of 5 to 0.01, were carefully prepared by mass balancing each component. \\
Figure \ref{fig: H calibration} shows the results of a benchmark calibration of a sample having roughly equal parts of hydrogen and aluminum (i.e. $H/Al\approx1$). All signals were acquired after carefully determining optimal excitation pulse lengths under sufficient spin relaxation-optimized pulse repetitions. Since for samples relatively rich in $^1H$, and signals detected in the hydrogen channel are rather intense and high $\zeta(^1H)$ values can be achieved after a fairly short acquisition time, we choose to fix $\zeta(^1H)$ at a predetermined point to observe the influence of the SNR ratio $\hat{\zeta}$ on the calculated relative hydrogen concentration $n(^1H)$ using eq.\ref{1}.\\
As can be seen, for $\hat{\zeta}$ less than unity, calculated hydrogen concentrations $n$ vary significantly exhibiting uncertainties as high as 8$\%$, or about $1.41$ wt\% of hydrogen in the current sample. At $\hat{\zeta}\gtrsim 3$, $n$ converges to stable values and all acquired hydrogen concentrations after $\hat{\zeta}\gtrsim 5$ were found to fall within $n\pm1\sigma$, at which point $u(n)$, the uncertainties of $n$, become as small as 1.5$\%$, or $1.37$ wt\% in this sample. We found that any further data acquisition to enhance $\hat{\zeta}$ will only result in a minor decrease in $u(n)$ while $n$ remains almost unchanged, which usually terminates the running quantification experiment.     
All other $LaH_3/Al$ samples were analyzed in the same manner, see supplementary materials (Figure 1), and similar conclusions were found. Comparing computed hydrogen contents $n$ from $\mu$Q-NMR measurements with initial mass-balance preparation shows agreeable similarities between both methods, see table \ref{tab I}. Small deviations in $\mu$Q-NMR derived values of $n$ likely originate dehydrogenation effects of $LaH_3$ within an air atmosphere. Nevertheless, we regard the similarity of both values of high enough quality to further develop the method. Similar conclusions were found in an earlier experiment by Meier et al. \cite{Meier2023} in \textit{in-situ} NMR experiments in diamond anvil cells.\\
Quantification experiments of the above mentioned sample classes allowed for very accurate benchmarking of $\mu$Q-NMR, by comparison with standard quantification methods, from about $\sim10^5$ to $\sim10^2$ wt-ppm of hydrogen or water content in the respective samples. Figure \ref{fig: H comparison} shows a direct comparison between NMR derived hydrogen concentrations and calibrated standard methods. As can be seen, NMR derived values are congruent to the values derived by standard methods within some wt-ppm. Slightly higher error bars for the DAC based quantifications originate from experimental peculiarities associated with \textit{in-situ} high-pressure NMR in DACs.  
\\
\begin{figure*}
\includegraphics[width=0.7\textwidth]{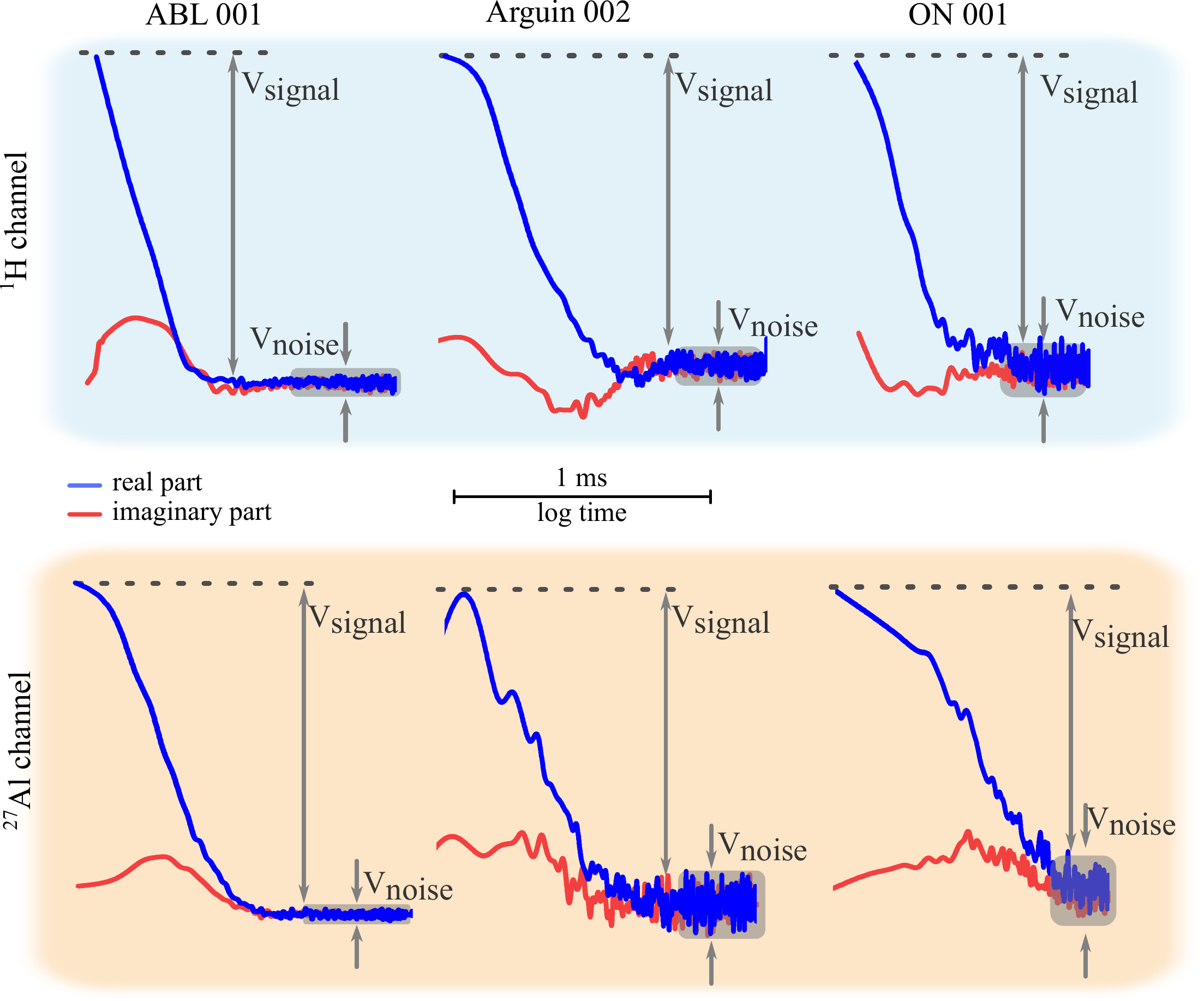}
\caption{\textbf{$^{27}Al$-NMR and $^1H$-NMR time-domain data acquired from three anorthites from meteorite ABL 001, Arguin 002, and ON 001 under ambient conditions.} The blue lines represent real parts and the red lines the imaginary parts of the induced signal amplitudes. The signal amplitude V$_{signal}$ in the time domain is taken at $t=0$; V$_{noise}$ was computed well with the range of stochastic thermal noise.}  \label{fig: time domains}
\end{figure*}
For further calibration, several natural pyolitic and rhyolitic glasses with varying hydrogen contents as well as nominally anhydrous mineral samples of Stishovite and Bridgmanite synthesised in multi-anvil analyses were used. The experimental procedure of determining both time-domain signal strengths, i.e. $\zeta(^1H)$ and $\zeta(^{27}Al)$, in each sample closely follows detection protocols described above. For a more detailed summary we like to refer the interested reader to the methods and materials section as well as to the supplementary material. The hydrogen/water contents of those samples were determined by either Nano-SIMS or FTIR analysis (see methods and materials section). As one can see both from table \ref{tab I} and figure \ref{fig: H comparison} the NMR derived hydrogen contents of those geologically important samples overlap well with values given by said standard methods.\\
To summarise, we were able to calibrate and benchmark $\mu$Q-NMR for hydrogen contents ranging over almost four orders of magnitude, down to the generally accepted detection limits of contemporary FTIR and Nano-SIMS analyses. 
\\
\begin{figure*}[htb]
\includegraphics[width=0.65\textwidth]{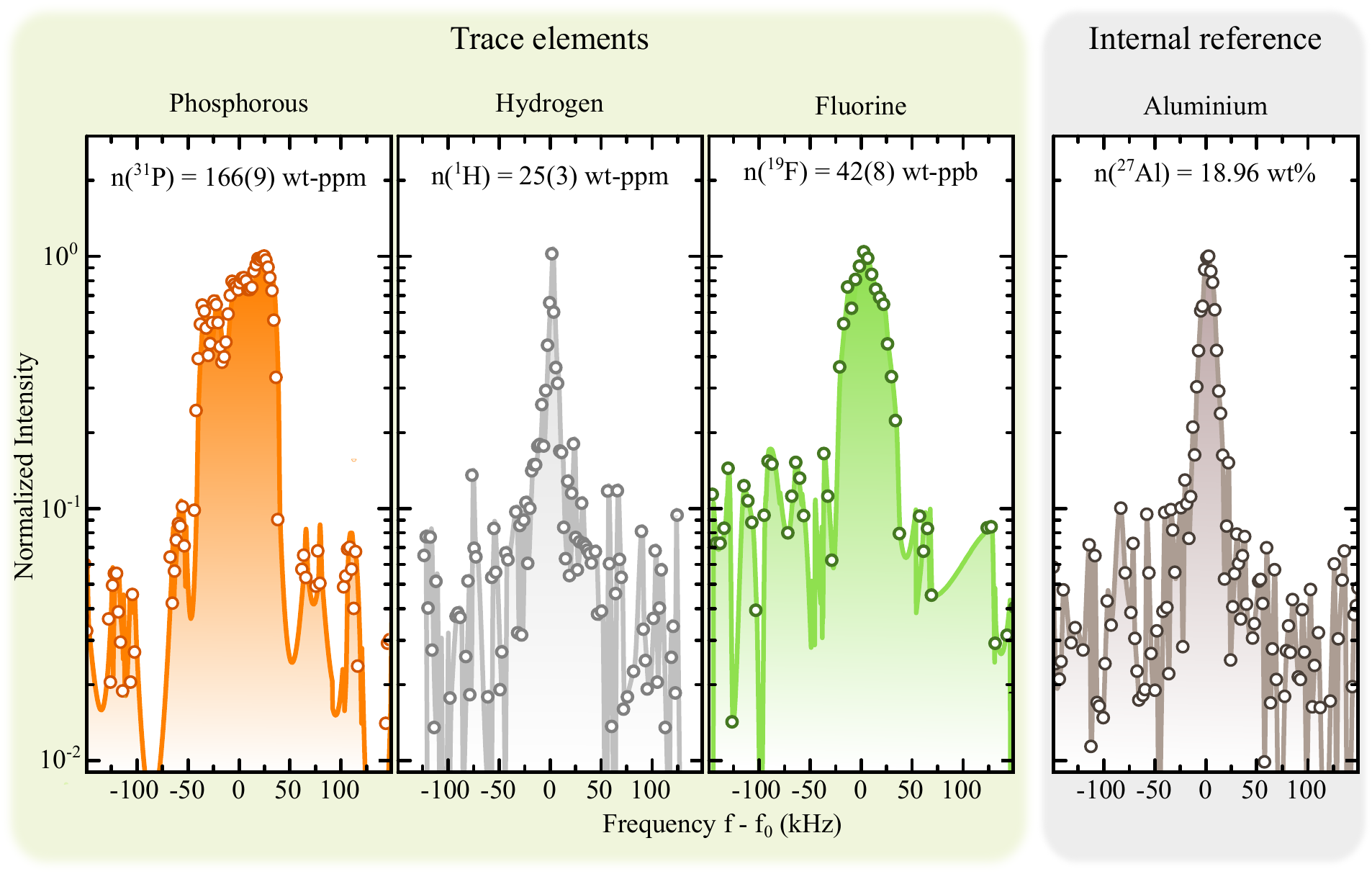}
\caption{\textbf{Representative single-grain multi-nuclear NMR spectra of meteoritic anorthosite Oued Namous 001}. Sample size was $\lesssim$ 5 nl (i.e. about $250$ $\mu m$ diameter and $150$ $\mu m$ height). All three trace elements (i.e. $^{31}P$, $^{1H}$ and $^{19F}$) exhibit reasonably well resolved resonance spectra after Fourier transform. Normalized spectra are plotted in logarithmic scales. All spectra were recorded after approximately $10^4$ scans. A detailed analysis of NMR spectra, apart from quantification will be published elsewhere.}  
\label{fig:ON1}
\end{figure*}   
\textbf{Application to meteorite-derived anorthites:}\\
Given limitations in reliably detecting trace element concentrations in the order of $10^{-1}$ - $10^1$ wt-ppm using FTIR or Nano-SIMS, no further calibration of our presented $\mu$Q-NMR approach was possible. However, Meier et al.\cite{Meier2017, Meier2018c} demonstrated that the time-domain limit of detection of Lenz-lens based micro-resonators can be as low as $10^{11} spins/\sqrt{Hz}$ which is about ten orders of magnitude more sensitive compared to static commercially available NMR equipment\cite{Meier2018b, Meier2018}. Provided that the here employed Lenz-lens based sample cavities are approximately three orders of magnitude bigger than those commonly used in diamond anvil cells \cite{Meier2018a}, we can gauge to first approximation, the time domain limit of detection of our resonators to be $\sim 10^{14}~spins/\sqrt{Hz}$. Provided digitization frequencies of $1$ MHz, our bigger resonators would be sensitive to $\sim 10^{17}$ spins within the sample chamber, which would correspond to detectable volumes of about $3~nl$ of $H_2O$.\\
Of course this back-of-the-envelope calculation assumes that no data averaging is applied and only single scans are recorded. Thus, considering data averaging, the detection limit might be enhanced by another two or three orders of magnitude. Further enhancement is commonly deemed unfeasible due to excessive demands on acquisition times. Therefore, we can assume that the mass-sensitivity of the here presented approach lies several orders of magnitude below standard quantification methods currently employed by the geo-science community.\\
In order to test this hypothesis we analysed three meteoritic plagioclases belonging to the anorthite category (see supplementary material Figure 2), namely Arguin002, Al-Bir Lahlou (ABL) 001 and Oued Namous (ON) 001. The backscatter electron image of those sample specimens were shown in Supplementary material Figure 3. All three polycrystalline samples were found to be a heterogeneous ensemble of anorthite, see supplemental material for a full analysis of their composition. Water concentrations in those nominally anhydrous samples are usually estimated to be below $50$ wt-ppm\cite{Sarafian2019, Hui2013, Sarafian2017, Newcombe2023a}.\\
Figure \ref{fig: time domains} shows the recorded raw data time domains for both $^{27}Al$- and $^1H$-NMR channels for all three meteoritic anorthites. We used single grain samples of approximately $100~\mu m$ to $250~\mu m$ in diameter to avoid inter-grain stoichiometric heterogeneities. Both signals in the hydrogen and aluminum channels were recorded with the same pulse sequences in order to keep acquired values of $\zeta$ as comparable as possible. For a detailed summary of the experimental parameters, see supplemental Table I. \\
The scan normalized signal-to-noise ratios $\zeta(^1H)$ where found to be $12.23,~0.15,~2.46$ for ABL001, Arguin002 and ON001 respectively. Considering the conversion from isolated hydrogen nuclei spins to molecular spins of $H_2O$, we found the three samples to contain $12(1)$ wt-ppm, $4(1)$ wt-ppm and $25(3)$ wt-ppm of molecular water, see supplemental Figure 4 and Table II, well within the expected ranges of water concentrations of meteoritic anorthites\ref{fig: water contents of NAMs} \cite{Newcombe2023a, Mills2017, Hui2017}.  
\\
\textbf{Wt-ppb $\mu$Q-NMR quantification of multi-species trace elements in a single mineral grain:}
\\

One of the major advantages of NMR spectroscopy is its spectral resolution and element specificity, fully preventing spectral overlap of signals stemming from different nuclear species, such as $^1H$ and $^{19}F$. This is often in gross contrast with other spectroscopic methods where molecular vibrations span wide spectral regions, grossly super-positioning with signals from other molecules or atoms. In particular, those issues make spectral interpretations in optical or vibrational spectroscopies very difficult and sometimes render a quantitative interpretation moot.\\
Thus, combining this unique feature of NMR spectroscopy with the here presented $\mu$Q-NMR approach, we decided to investigate other potential trace element concentrations in the angritic meteorite Oued Namous 001, which, besides trace amounts of hydrogen and water, is also believed to incorporate trace elements of phosphorous of $\sim150$ wt-ppm and minor traces of fluorine of $\sim100$ wt-ppb\cite{Newcombe2023a, Sarafian2019}.
\\
Detecting both $^{31}P$ and $^{19}F$ in ON 001 places two special experimental challenges. On the one hand, phosphorous is not, unlike $^1H$ and $^{19}F$, a "high $\gamma_n$" nucleus, leading to significantly reduced signal amplitudes per scan. On the other hand, while fluorine exhibits one of the largest gyromagnetic ratios after hydrogen, the extremely low anticipated concentration levels of $\sim10^2$ wt-ppb will lead to very low signal intensities as well.
\\
$^{19}F$ and $^{31}P$-NMR spectra of ON 001 were acquired using the same resonator used to determine hydrogen concentrations in the same sample specimen. Both spectra were recorded after $\gtrsim 10^4$ accumulated scans to ensure sufficient signal-to-noise ratios $\zeta$ in both channels. Supplementary Table III provides a summary of the experimental parameters during those experiments. All NMR spectra were recorded using solid echoes with a gaussian RF pulse modulation. We found that this modified spin echo represents an optimal trade-off between excitation band width as well as distortion- and background-free spectra. Figure \ref{fig:ON1} shows four representative NMR spectra of the to-be-quantified nuclear spin subsystems of the trace elements $^1H$, $^{19}F$ and $^{31}P$ as well as of the internal aluminum reference. All four spectra displayed reasonably sharp lineshapes, with $^1H$ and $^{27}Al$ linewidths being about fourfold than those of fluorine and phosphorous. For the time domain data, see supplemental material Figure 5.\\   
Employing the same strategy as outlined before - i.e. ensuring full spin excitation by conducting RF-pulse nutations for every spin sub-system as well as carefully determining spin lattice relaxation times in order to ensure spectra shown in figure \ref{fig:ON1} were being recorded after full equilibration - we found the relative phosphorous concentration $\hat{n}(^{31}P)$ to be $7.7(3)\cdot 10^{-4}~spins/^{27}Al$ and  $\hat{n}(^{19}F)=3.2(4)\cdot 10^{-7}~spins/^{27}Al$ for the fluorine sub-system. Using micro-probe analysis, we found that a global bulk concentration of $18.96~wt\%$ of $^{27}Al$ is present in our sample, see supplemental materials. Therefore, the actual trace element concentrations of both phosphorous and fluorine were computed to be: $n(^{31}P)=166(9)$ wt-ppm and $n(^{19}F)=42(8)$ wt-ppb. These findings are in very good agreement with first-hand estimates.  
\section{Conclusions}

We have presented a novel approach for trace-element detection in geo-physically important minerals using internally self-referenced and self-consistent quantitative NMR employing state-of-the-art magnetic flux tailoring techniques. We christen this method "$\mu$Q-NMR" for its ability to quantify volatiles in micro-meter sized single individual grains. We could show that $\mu$Q-NMR yields mass-sensitivities in excellent agreement with contemporary quantification methods used in geo-science and comparable disciplines spanning over four orders of magnitude. Our showcase examples, meteorite-derived anorthites, demonstrated that $\mu$Q-NMR can routinely be used to detect hydrogen concentrations in the order of 10 wt-ppm without the need for calibrations or cross-referencing. Furthermore, we demonstrated that $\mu$Q-NMR is not limited to quantification of hydrogen or water concentrations but can also be used to constrain other trace element amounts like phosphorous or fluorine. Finally, we were able to show that our presented method is capable of detecting concentrations on wt-ppb level of high-$\gamma_n$ nuclei.
\\
The reported $^{19}F$ concentration of $42(8)$ wt-ppb represents an enhancement of about two to three orders of magnitude compared to commonly achievable detection limits reported in FTIR or even Nano-SIMS\cite{Saal2008}. It should be noted, that our finding is not an \textit{a-priori} limit of detection. The actual detection limit of $\mu$Q-NMR is closely related to the resonator size, i.e. sensitivity $\propto V^{-1/2}$. If we follow the inverse argument given in the last chapter and assume that our reported fluorine concentration represents a real sensitivity limit, then by reducing the volume of the resonator, this limit would be about $20$ wt-ppb for a volume half the size of our currently used set-up; and one-tenth of our reported value for resonator cavities as small as $50$ $\mu m$ in diameter and height. In the same vein, increased accumulation times will also somewhat improve detection limits, but as pointed out earlier, this "brute-force" method is limited by excessive time demands on the NMR spectrometer consoles. 
\begin{figure*}[htb]
\includegraphics[width=0.8\textwidth]{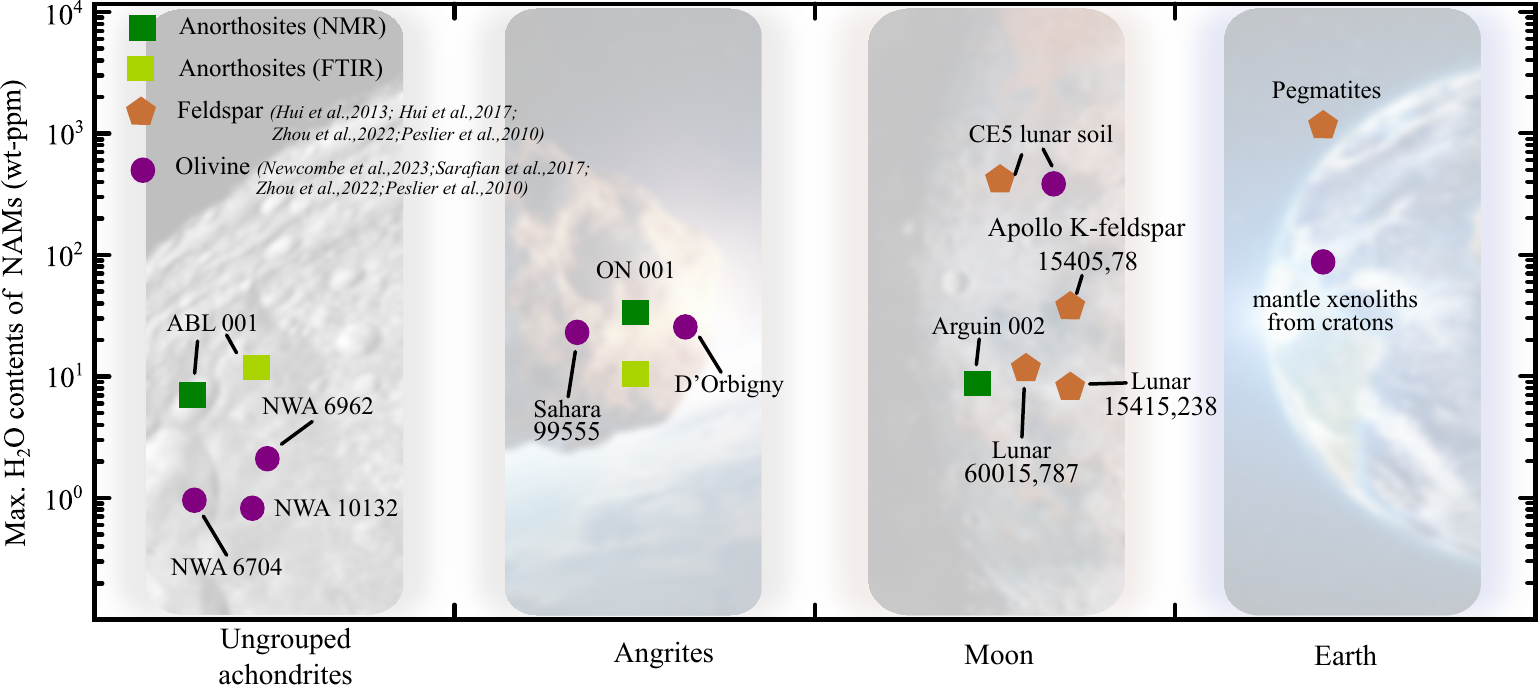}
\caption{\textbf{Maximum water contents of anorthites (this study) and other NAMs from ungrouped achondrite meteorites\cite{Newcombe2023a}, angrites\cite{Sarafian2017}, the Moon\cite{Hui2017, Hui2013, Mills2017, Zhou2022} and the Earth\cite{Peslier2010, Johnson2004} }. This figure is modified from Newcombe et al.,2023\cite{Newcombe2023a}. Anorthites of ABL 001 and ON 001 were simultaneously analyzed by FTIR, see supplementary material Figure 6 and Figure 7. Three ungrouped achondrite olivines are from three individual parent bodies. Two lunar feldspars ( K-feldspar and anorthites) from ferroan anorthite and Mg-suite troctolite were analyzed by Mills et al.,2017 and Hui et al.,2017\cite{Mills2017, Hui2017}and the feldspar and olivine from CE5 lunar soil were reported by Zhou et al.,2022\cite{Zhou2022}.}  
\label{fig: water contents of NAMs}
\end{figure*}   
\\
Despite its obvious advantages, $\mu$Q-NMR is subject to its own limitations. Firstly, it was implicitly taken for granted that all experiments are done in the same resonator using the same RF-coil in a single channel home-build NMR probe. The special design of our probes allows for free tuning and matching from frequencies of about $20$ MHz to the highest frequencies of hydrogen without significantly altering any RF routing of the probe except for small adjustments on the tuning path. Usual commercially available single channel probes are set-up to either work at hydrogen frequencies at specific magnetic fields or for a rather limited "X-channel" covering frequencies of most low-$\gamma_n$ nuclei. A possible work around of this issue might be the use of double channel NMR probes working simultaneously at hydrogen and X-channel frequencies. These probes are usually based on the principle of coupled RF tank circuits, with two separate circuits detecting for example $^1H$ and $^{27}Al$ being connected at the NMR coils interfaces. Since our presented $\mu$Q-NMR approach relies on the fact that the same tank circuit is used for both nuclei channels, coupled double channel NMR probes might exhibit lower mass-sensitivities. Therefore, usage of specialized home-built NMR probes of high quality is mandatory. 
\\
Albeit being a seemingly simple approach, a successful quantification experiment also greatly relies on sample quality and local atomic environments. This is particularly true if the internal reference nuclear spin sub-system is quadrupolar, i.e. the spins have a nuclear spin greater than 1/2, such as $^{27}Al$. In those cases it is vital to achieve a full spin excitation of the system in order to achieve maximum values of $\zeta(^{27}Al)$ and compute reliable trace element concentrations. Alumina ($Al_2O_3$) in particular is known to exhibit very broad resonance spectra in excess of 800 kHz or more, depending on the local charge symmetry around the aluminum nucleus. In case of low local symmetry, resonance spectra might become too broad for sufficient RF  spin excitation, losing its value as an internal reference standard. All of the samples investigated in this work exhibited very sharp $^{27}Al$ resonance spectra, indicating fairly symmetric local environments.
\\
Another common issue using $\mu$Q-NMR is the presence hydrogen backgrounds and spurious signals stemming from organic adhesives or wire insulations of the excitation coils. This problem can generally be avoided by careful assembly of the Lenz-lens based resonators and conservative RF powers used for excitation. For NMR resonator volumes as small as the ones used in this work, RF powers of about 1W or less are sufficient to achieve optimal spin excitation of the sample in the resonators center\cite{Meier2018c}. Therefore, any hydrogen reservoirs located at outer regions of the resonator or the driving coils will not be sufficiently excited to contribute to any excessive hydrogen backgrounds. Throughout all successful experiments presented here, no significant hydrogen backgrounds were recorded.
\\
The non-destructive and independent analysis of $\mu$Q-NMR is in notable contrast to previous measurements for quantifying water contents of rare materials. It is compatible with other referenced materials from  $\sim10^5$ to $\sim10^2$ wt-ppm and deficient volatile concentrations of meteorite-derived anorthites. Due to the complex petrogenesis and Monotonic meteorites, we do not attempt to calculate a bulk parent body $H_2O$ content using petrogenetic models. Nevertheless, comparing water concentrations among NAMs of ungrouped achondrite meteorites, Angrites, lunar, and terrestrial origin, there is at least an order of magnitude relative to the Earth in Figure \ref{fig: water contents of NAMs}. In other words, this finding implies that these materials may not be major sources of Earth’s $H_2O$.
\\
Any NMR-active isotope with a non-zero spin quantum number can be quantified and cross-verified under ideal conditions. With respect to volatiles, controversial systems of terrestrial and extraterrestrial origin exhibiting uncertain water contents could be re-evaluated. Future work focusing on quantitative analysis of representative materials with extremely volatile deficits as well as the qualitative judgment of volatile (i.e. \textit{H},\textit{F},\textit{Cl},\textit{S}) will be needed in the future to put constraints on the time and spatial evolution process of distribution of volatile compounds over the formation and accretion history of Earth and other planets.
\\
Finally it should be noted that the here presented approach is not limited to geological samples. In fact, initial quantification experiments by our group have been carried out on potential high-temperature superconductors and other correlated electron systems. However, this method is particularly attractive for the geo-science community due to its extremely low detection capabilities of NMR active nuclei, its calibration-free nature and self-consistency as well as its non-destructiveness. The last point in particular makes $\mu$Q-NMR attractive when extremely precious sample specimens need to be investigated. Therefore, the here presented method could play a revolutionizing role in - but not limited to - contemporary geo-chemistry and -physics. 

\section{Materials and Methods}

\subsection{Materials}
Seventeen samples were used as reference materials for calibrating water or hydrogen concentrations using the $\mu$Q-NMR approach presented in this work, including rhyolitic and pyrolitic glasses of natural and synthetic origin, several metal hydride mixtures, and two nominally anhydrous minerals (stishovite, bridgmanite) from multi-anvil syntheses. Additionally, we analyzed three nominally anhydrous anorthites of meteoritic origin. Supplementary material Table IV summarises all samples' bulk composition. 
\\
\textbf{\textit{Rhyolitic and pyrolitic glasses}}:
The composition of used glasses, natural rhyolitic obsidian (OA-1), synthetic rhyolitic hydrous glasses (RH-4) and (RH-9), and pyrolite glasses PYR-01 to PYR-03 were determined by electron probe microanalysis (EPMA), detailed syntheses were published elsewhere\cite{Tu2023, Wu2022}. Rhyolitic obsidian OA-1 was analyzed for its hydrogen concentration (expressed as molecular water) by FTIR and Nano-SIMS, yielding $1214(30)$ wt-ppm and $1149(87)$ wt-ppm of water present in those samples respectively. Water contents of synthetic rhyolitic glasses RH-4 and RH-9 were obtained by a thermal conversion elemental analyzer with isotope ratio mass spectrometry (TC/EA-MS), Fourier transform infrared spectroscopy (FTIR), and confocal Raman spectroscopy.
Average water contents of (RH-4) and (RH-9) were $1.38~wt\%$ and $5.66~wt\%$ respectively\cite{Tu2023}. The pyrolitic composition was found iron-free. Samples have been synthesized under an $8~\%~H_2$ gas flux. Water contents were found to be $700(100)$ wt-ppm, and $900(200)$ wt-ppm respectively. Sample sizes varied from $1~mm$ to $3~mm$ in diameter, were thoroughly cleaned using anhydrous ethanol and dried for more than $24~h$ to remove surface water. 
\\
\textbf{\textit{Metal hydride mixtures}}:
For a more systematic calibration of the presented method, we used pre-defined mixtures of $LaH_3$ and $Al$ powder (Sigma Aldrich, $4$N purity). Nine samples with hydrogen-to-aluminum ratios spanning two orders of magnitude from $0.01$ to $5$  were mixed by carefully weighing individual compounds using a high-precision balance. All samples were sealed in PTFE tubes to avoid hydrogen contamination and oxidation.
\\
\textbf{\textit{Nominally Anhydrous Minerals}}: 
Small samples, typically less than $100~\mu m$ in diameter, of bridgmanite and stishovite were analyzed for their water contents using Nano-SIMS, constraining their water contents to $1100(150)$ wt-ppm and $6200(1500)$ wt-ppm \cite{Li2023} respectively. Additionally, we analyzed three anorthites from meteorite specimens, namely, Al-Bir Lahlou 001 (ABL 001), Arguin 002, and Oued Namous 001 (ON 001). Typical grain sizes were observed to be around $250~\mu m$ in diameter. Most meteorites containing feldspar are highly internally fractured, resulting in challenging analyses of volatile elements\cite{Sarafian2019}. These samples were selected due to a high abundance of feldspar. In the following, we summarise the petrological descriptions of these three anhydrous anorthites.
\\
Arguin $002$ is a lunar norite predominantly composed of orthopyroxene ($54.2\%$) and plagioclase ($39.1\%$), with a small amount of clinopyroxene ($4.1\%$) and silica ($2.5\%$)
(Supplementary material Figure 3A). Trace phases in the meteorite include chromite, ilmenite, troilite, and zircon. Orthopyroxene and plagioclase occur as a mixture of large fractured grains, compositionally consistent megacrysts, with sizes reaching up to $3~mm$. Clinopyroxenes occur exclusively as exsolution patches or elongated grains within the host orthopyroxene, measuring typically $5-10~\mu m$ in width. In addition, impact melt glass is ubiquitously dispersed along inter-crystal boundaries and intra-crystal fractures, having a composition consistent with mixtures of plagioclase and clinopyroxene. Other minor minerals are sporadically found in the interstitial regions or within the impact glass. 
\\
Al Bir Lahlou (ABL) $001$ is a norite mainly consisting of clasts of anorthite and low-Ca pyroxene, which occur as a mixture of large fractured grains (~a few $mm$)(Supplementary material Figure 3B), or as fine-grained crystals interstitial to the coarser grains. Minor phases include olivine, ilmenite, and chromohercynite. Olivine is fairly rare, and only a few grains with sizes ranging from  $30-150~\mu m$ were observed as inclusions in pyroxene. The composition of all measured minerals is relatively homogeneous. Even if the meteorite most likely originated from Vesta, its oxygen isotope values and An values deviate from typical HEDs values\cite{Gattacceca2022}.
\\
Oued Namous (ON) $001$ is a volcanic angrite based on the significant igneous zoning of olivine and pyroxene. One section of the meteorite had been observed on scanning electron microscopy (SEM) for petrography in this study(Supplementary material Figure 3C and Figure 3D). The fayalite, kirschsteinite, \textit{Al}-\textit{Ti}-rich clinopyroxene, and anorthite were the major components that sum of abundance covered $90$\% in this section. It also contains a large number of plagioclases, and the diameter of the plagioclase in the angrite analyzed in this study is about $2~mm$. Smaller sample sizes of about $300~\mu m$ were selected after careful factorization\cite{Gattacceca2023}. 
\subsection{Methods}
\subsubsection{NMR measurements}
Nuclear Magnetic Resonance experiments were carried out on a $Bruker~400$ magnet (AVANCE III NEO) and a retrofitted $Oxford~400$ magnet at the Center for High-Pressure Science \& Technology Advanced Research, Beijing. Spectra were recorded at a static magnetic field of about $9.4~T$, corresponding to $400.2$ MHz and $104.3$ MHz for $^1H$ and  $^{27}Al$ resonance frequencies respectively. The corresponding frequencies for $^{27}Al$ ,$^{19}F$ and $^{31}P$ were $104.227$ MHz, $376.376$ MHz and $161.923$ MHz under the $9.278~T$ of the retrofitted Oxford magnet. We used a fully home-built NMR probe customized to the special demands for detecting faint nuclear induction signals of samples of less than $50~nl$ (roughly $400~\mu m~\times~400~\mu m~\times~400~\mu m$). To ensure proper spin excitation and maximized signal amplitudes, pulse nutations were carried out for each sample and respective nucleus channel. All spectra were recorded after allowing for sufficiently long spin relaxations.
\\
We conducted experiments at similar conditions before and after sample loading to eliminate the possibility of spurious signals (in particular occurring in the $^1H$ channel). No additional $^1H$-NMR signals were found, thus we can expect all recorded free-induction decays or spin echoes to originate from the samples alone. Signal-to-noise ratios were acquired in the time domain. 

\subsubsection{Scanning Electron Microscopy and Electron Probe Microanalysis}

 Petrographic observations for meteorites were carried out using the JSM-7900F field emission scanning electron microscope at the HPSTAR, using electron beam currents of $2.0–6.4~nA$ and an acceleration voltage of $15~kV$.The major element compositions of natural hydrous and anhydrous silicate minerals were obtained by JEOL JXA-8530F Plus Field Emission Electron Probe at the University of Science and Technology of China (USTC) and JEOL JXA-8230 electron microprobe at the Chinese Academy of Geological Sciences (CAGC). The  $15~kV$ accelerating voltage, $20~nA$ beam current, and $2–10~\mu m$ beam diameter were set up to analyze minerals.
\\
 Natural minerals and synthetic oxides were used as standards, and the quality of analyses was assessed based on stoichiometric constraints. Elements chosen to be determined were calibrated by the program based on the ZAF procedure. The stoichiometry and ferric iron content of minerals were calculated based on charge balance. Individual grains were found to have homogeneous major and minor element contents.
  
\subsubsection{Nanoscale Secondary Ion Mass Spectroscopy}

The $H_2O$ contents of rhyolitic glasses (RH-4 and OA-1) and anorthites from meteorites ON 001 and Arguin 002 were measured using NanoSIMS under dynamic mode at the State Key Laboratory of Isotope Geochemistry, Guangzhou Institute of Geochemistry, Chinese Academy of Sciences, Guangzhou. Measurements were conducted using the CAMECA NanoSIMS $50$L scanning ion microprobe following the method by Yang et al., 2023\cite{Yang2023}. A primary beam of Cs$^+$ was used to raster the target sample area and electron multipliers were used to collect negative secondary ions $^{12}$C$^-$, $^{16}$OH$^-$ and $^{30}$Si$^-$ simultaneously.The nuclear magnetic resonance probe was activated along with the Hall probe to correct potential magnetic field drift and ensure a stable magnetic field during the analysis. 
\\
 Before each analysis, a primary beam of $3~nA$ was used to sputter the target area of $ 35~\mu m$ × $35~\mu m$ for $3$ minutes followed by a primary beam of $0.6~nA$ was used to raster the central $30~\mu m$ × $30~\mu m$. With electronic blanking applied, only signals from the central $10~\mu m$ × $10~\mu m$ area were recorded. Fifty cycles of data with $1$ second for each cycle were collected. The data were corrected for EM deadtime ($44~ns$). $^{30}$Si$^-$ normalized secondary ion ratios of basaltic glass standards MRG-G1\cite{Shimizu2017}  were used to derive sensitivity factors to calibrate $H_2O$ contents. The background level of $H_2O$ monitored using Suprasil $3002$ glass \cite{Yang2023} is about $2.8$ wt-ppm. The detection limit is about $3.5$ wt-ppm, which is derived from background plus $3$ times the standard deviation of background.

 \subsubsection{Fourier Transform Infrared Spectroscopy}
 
 Anorthites of meteorite origin (ABL 001 and ON 001) were selected under an optical microscope and then embedded in tin-bismuth low-temperature alloy for double-sided polish, to obtain crystal plates with the thickness of $200-300~\mu m$\cite{Zhang2018}.FTIR spectroscopy measurements were carried out on crystal plates using a $Thermo~Scientific~Nicolet~iS50$ spectrometer at the Institute of Geochemistry of the Chinese Academy of Sciences (IGCAS). The spectrometer equipped with an MCT detector ($11700-600$ $cm^{-1}$) and DTGS ($7800-350$ $cm^{-1}$), spectral resolution better than $0.1$ $cm^{-1}$, sensitivity better than $55000:1$. The spectra of anorthites were acquired in the range from $6000-650$ $cm^{-1}$, the light beam of $75$×$75~\mu m$, scan number of $256$, the spectral resolution of $4$ $cm^{-1}$ and all tests conducted at nitrogen background and room temperature. The water content of tested samples ($C_{H_2O}$ in wt-ppm) could be derived from the molar absorption coefficient (I in $ppm^{-1}.cm^{-2}$) of $17.3(1.1)$\cite{Mosenfelder2015,Johnson2003}, sample thickness (D in $cm$) and the $3700-3100$ $cm^{-1}$ integrated area (A of $cm^{-2}$) which correspond to hydroxyl vibration, according to the Beer-Lambert law for the equation of $C_{H_2O}$=$3$A/(I×D)\cite{Hui2013,Zeng2021}, the uncertainty of calculated water contents was estimated to be $30\%$.
 
\section*{Acknowledgements}
This work was financially supported by the National Natural Science Foundation of China (grant No. 42150104 to R. Tao) and the National Key Research and Development Program of China (2019YFA0708501 to L. Zhang). We thank Prof. Xiaoying Gao from the University of Science and Technology of China (USTC) for providing glass samples (RH-4 and RH-9) and Dr. Wancai Li from the USTC for help with EPMA.

\section*{Author Contributions Statement}
Conceptualization: R.T., T.M., S.L. and L.Z.; Methodology: Y.F., T.M. and R.T.; Investigation: Y.F., T.M., Y.Y., D.S. Z.W. and R.T.; Visualization: Y.F., and T.M.; Formal analysis: Y.F., T.M., and R.T.; Validation: Y.F., T.M., and R.T.; Funding acquisition: R.T. and L.Z.; Resources: R.T., T.M. and L.Z.; Project administration: T.M. and R.T.; Writing—original draft: Y.F.; Writing—review and editing: Y.F., T.M. and R.T.

\section*{Competing Interests Statement}
The authors declare that they have no known competing financial interests or personal relationships.

\section*{Data availability}
Data will be made available on request.


\end{document}


\title{Supplementary material}

\date{\today}
\maketitle             

\FloatBarrier
\section{Table}

\begin{table}[h!]
\centering
 \caption{Experimental parameters of Micro-NMR measurements for anorthites in meteorites}
    \begin{tabular}{cccc}
         \hline
Parameters/Sample & ABL001 & Arguin 002& ON 001 
           \\
           \hline
Scans $^1H$ & 10480 & 51200 &10240
\\
SNR$^1H$ & 1252.51 & 33.63 & 249.14
\\
 Dw $^1H$/$\mu s$ & 2 & 1.3 &2
 \\
 $\zeta_H$ & 12.23& 0.15 &2.46
 \\
 Scans $^{27}Al$ & 5120 & 2048 &512
 \\
 SNR $^{27}Al$ & 1573.37 & 16.17 &111.25
 \\
 Dw $^{27}Al$ & 0.5 & 1 &0.5
 \\
$\zeta_{Al}$& 21.98 & 0.36 & 4.91
\\
n$^{27}Al$/at\%  & 14.92 & 14.31 &14.91
\\
$n_H/n_{Al}$/$\times10^{-4}$  & $1.89(0.11)$ & $0.55(0.03)$ &$3.03(0.20)$
\\
$n_{H_2O}$/wt-ppm  & $11.64 (0.67)$ & $4.41 (0.24) $ &  $24.81(1.63)$
\\
         \hline
    \end{tabular}
    \label{tabI}       
   \end{table}
 
\begin{table}[h!]
\centering
\caption{water contents of anorthites (this study) and other NAMs from ungrouped achondrite meteorites\cite{Newcombe2023a}, angrites\cite{Sarafian2017}, the Moon\cite{Hui2017, Hui2013, Mills2017} and the Earth\cite{Peslier2010, Johnson2004} }
\resizebox{1.0\textwidth} {!}{
    \begin{tabular}{cccccccccc}
         \hline
\multirow{3}{*}{Planet} & \multicolumn{3}{c}{\makecell{anorthite(this study)}}&\multicolumn{3}{c}{\makecell{anorthite(previous study)}}& \multicolumn{3}{c}{olivine}
\\
     \cline{2-10}
& meteorite &\makecell{$H_2O$ content\\/wt-ppm}&Unc.&meteorite&\makecell{$H_2O$ content\\/wt-ppm}&Unc.&meteorite&\makecell{$H_2O$ content\\/wt-ppm}&Unc.
\\
 \hline
\multirow{2}{*}{\makecell{Ungrouped\\ achondrite}} &\multirow{2}{*}{ABL 001}&\multirow{2}{*}{11.64}&\multirow{2}{*}{0.67}&\multirow{2}{*}{ABL 001}&\multirow{2}{*}{17.74}&\multirow{2}{*}{2.22}& NWA6704&0.88&0.13
\\
&&&&&&& NWA6962&1&0.38
\\
\hline
\multirow{2}{*}{Angrites}&\multirow{2}{*}{ON 001}&\multirow{2}{*}{24.81}&\multirow{2}{*}{1.63}&\multirow{2}{*}{ON 001}&\multirow{2}{*}{-}&\multirow{2}{*}{-}& Sahara 99555 &13 &6
\\
&&&&&&& D’Orbigny&14.1245&3
\\
\hline
\multirow{5}{*}{Moon}&\multirow{2}{*}{Arguin 002}&\multirow{2}{*}{4.41}&\multirow{2}{*}{0.04}&Lunar 60015,787&5&1&\multicolumn{3}{c}{\multirow{2}{*}{Nope}}
           \\
           &&&&Lunar 15415,238&4&1&&&
           \\
 &\multirow{3}{*}{CE5}&\multirow{3}{*}{-}&\multirow{3}{*}{-}&CE-PL 1&385&27& CE-OL 1&311&30
 \\
 &&&& CE-PL 2&231&16&CE-OL 2&165&16
 \\
 &&&& CE-PL 3&269&19&CE-OL 3&152&14
           \\
           \hline
 Earth&  \multicolumn{3}{c}{Nope}   &   \multicolumn{3}{c}{0-1035}& \multicolumn{3}{c}{0-80}
 \\
         \hline
    \end{tabular}
    }
    \label{tabII}       
   \end{table}
  
\clearpage
\begin{table}[h!]
\centering
 \caption{Experimental parameters of $^{27}Al-NMR$,$^{19}F-NMR$ and $^{31}P-NMR$ measurement about anorthite from angrite ON 001}
    \begin{tabular}{cccc}
         \hline
Parameters & $^{27}Al$ & $^{19}F$&$^{31}P$ 
           \\
           \hline
$t_{2/\pi}/\mu s$ & 12 & 6 &8
\\
I & 2.5 & 0.5 & 0.5
\\
f/MHz & 104.227 & 376.376 &161.923
 \\
 Dw/$\mu s$ & 1& 0.5 &1
 \\
 $\gamma/(\times10^{7}~rad/Ts$) & 6.976 & 25.181 &10.839
 \\
 Q & 90 & 100 &139.13
 \\
 Ns &10240 & 25600 &10240
 \\
$\zeta=SNR/\sqrt{Ns}$& 0.2933 & 0.0865 &0.0252
\\
$n_{mea.}/n_{{^{27}Al}}$  & 1 & $3.17\times10^{-7}$ & $7.70\times10^{-4}$
\\
abundance & 18.78~wt\% & $42(8)$ wt-ppb & $166(9)$ wt-ppm
\\
         \hline
    \end{tabular}
    \label{tabIII}       
   \end{table}
 \begin{table*}[h!]
\centering
 \caption{Major compositions of materials analyzed by EPMA in this study} 
    \begin{tabular}{ccccccccccccccc}
         \hline
Sample & $SiO_2$ & $TiO_2$ & $Al_2O_3$ &$FeO(T)$ & $MnO$ & $MgO$ & $CaO$ & $Na_2O$ & $K_2O$ & $Cr_2O_3$& $BaO$ & Total &Ref.
         \\
         \hline
OA-1 & 76.30 & 0.09 & 13.00 & 0.56 & 0.07 & 0.06 & 0.51 & 4.11 & 4.58 & -& -& 99.28 &\cite{Wu2022}
           \\
RH-4 & 74.69 & 0.13 & 11.15 &2.56 & 0.09 & - & 0.17 & 4.31 & 4.31 & 0.01& -& 97.41 &\cite{Tu2023}
            \\
RH-9 &73.41 & 0.12 & 10.92 & 2.12 & 0.08 & - & 0.17& 2.43 & 4.06 & 0.01 & - & 93.32 &\cite{Tu2023}       
            \\
PYR-01 & 50.71 & - & 4.36 &- & - & 42.53 & 2.40 & - & - & - & - &100&-  
            \\
PYR-02 & 50.71 &- & 4.36 &- & - & 42.53 & 2.40 & - & - & - & -&100&- 
            \\
PYR-03 & 50.71 & - & 4.36 &- & - & 42.53 & 2.40 & - & - & - & - &100&- 
            \\
An1.& 42.81 & 0.01 & 33.42 & 0.15 & 0.01 & 0.11 & 19.56 & 0.52 & 0.07 & 0.01  & -& 96.67 &-
            \\
An2. & 45.67 & 0.02 & 33.78 &0.13 & 0.01 & 0.08 & 17.92 & 0.75 & 0.09 & 0.01 & -& 98.46 &-
            \\
An3. & 43.69 & 0.01 & 35.37 &0.26 & 0.01 & 0.08 & 20.31 & 0.01 & 0.01 & 0.01 & -& 99.76&-
           \\
         \hline
    \end{tabular}

An1.- anorthite from ABL 001, An2.- anorthite from Arguin 002, An3.- anorthite from  Oued Namous 001. 
 Samples PYR-01 to PYR-03 were made in an 8\% $H_2$ gas flux.
    \label{Table IV}       
   \end{table*}
\FloatBarrier

\clearpage
\section{Figure}
\begin{figure}[h]
\includegraphics[width=0.8\columnwidth]{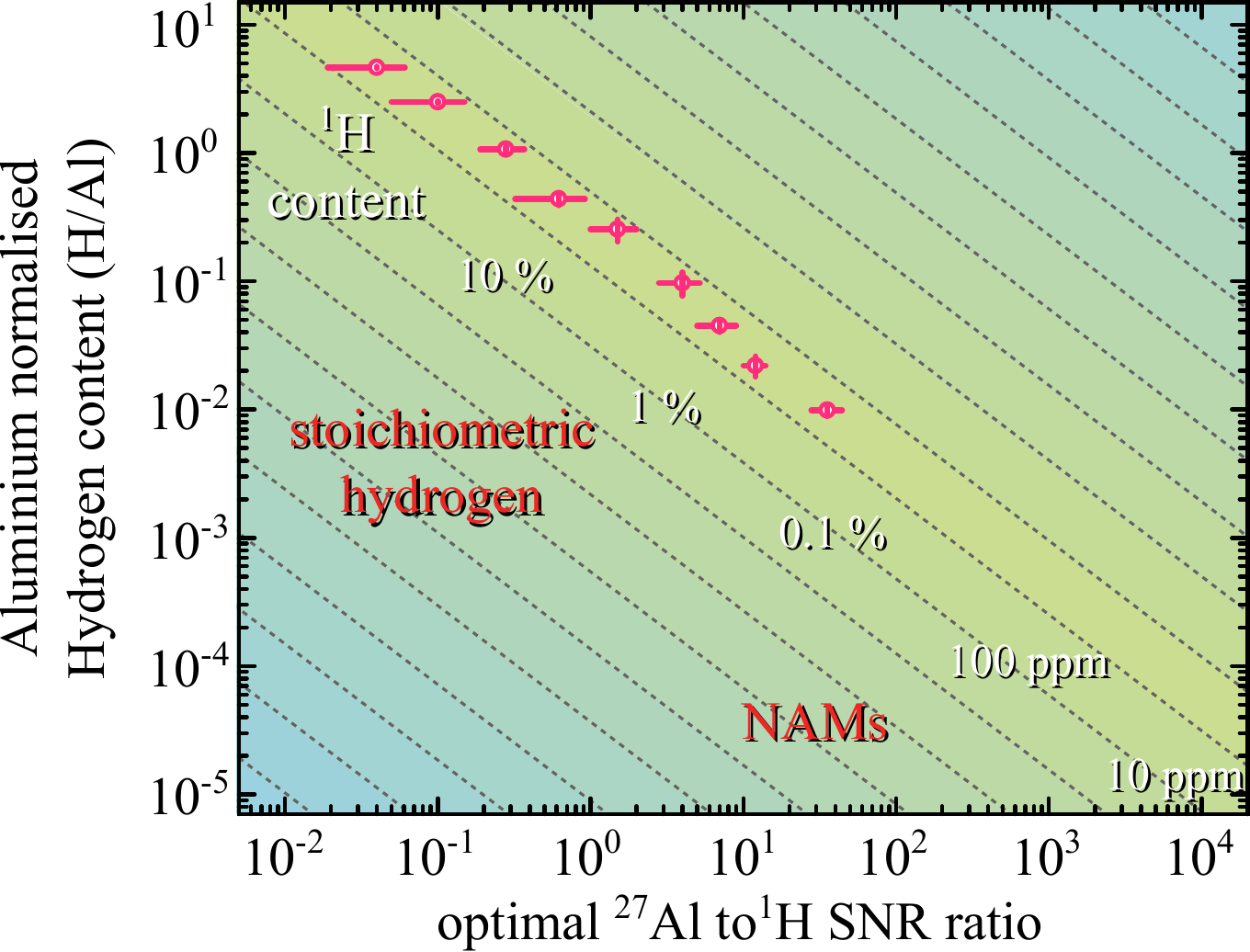}
\caption{\textbf{the relationship between Aluminium normalised hydrogen content and optimal $^{27}Al$ to $^1H$ SNR ratio}}
\label{fig.H calibrations LaH3 and Al  }
\end{figure}
\begin{figure*}[htb]
\includegraphics[width=0.8\textwidth]{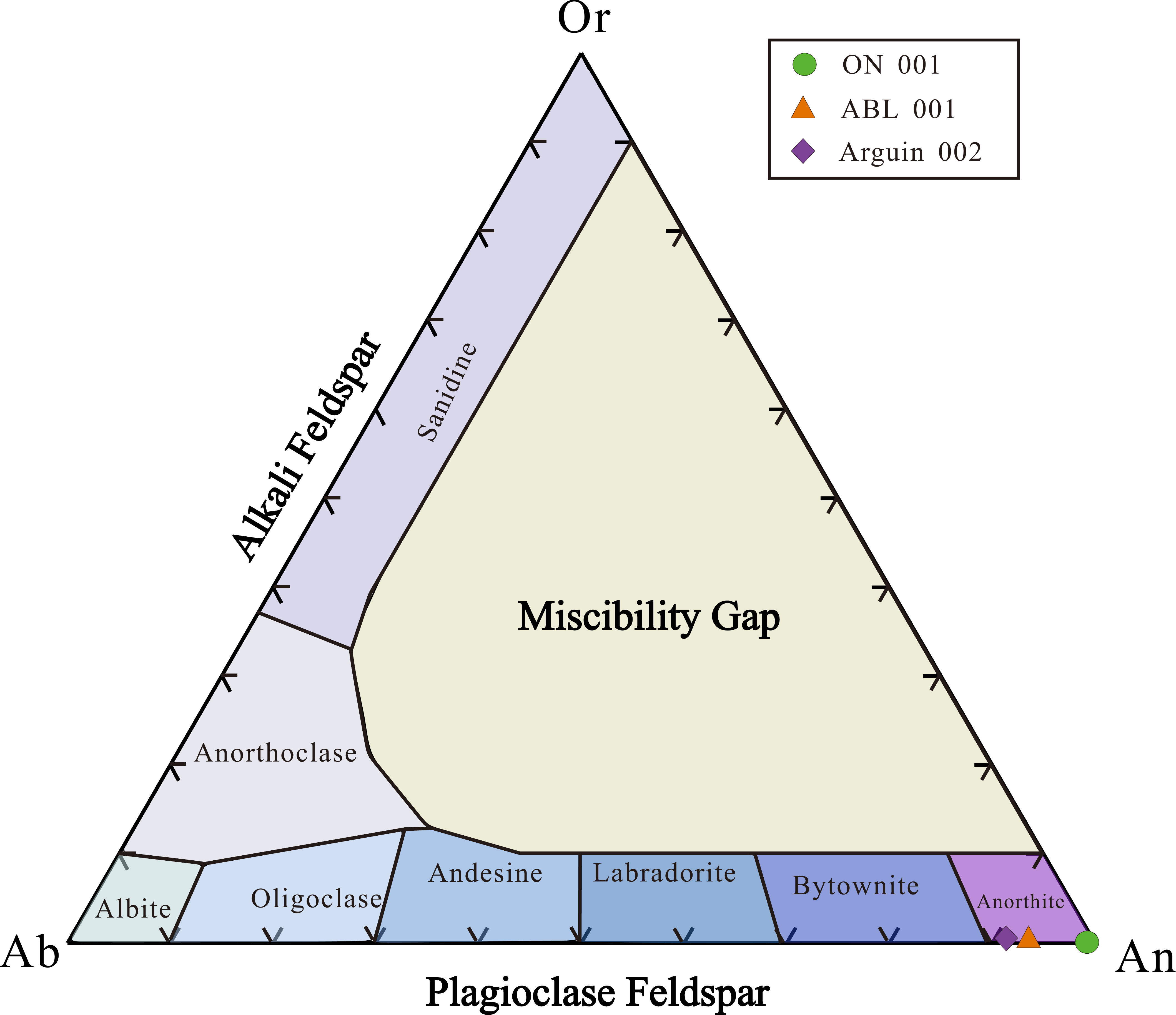}
\caption{\textbf{plot diagram of Plagioclase originating meteorites.} }
\label{fig: An plot}
\end{figure*} 
\begin{figure*}[htb]
\includegraphics[width=0.8\textwidth]{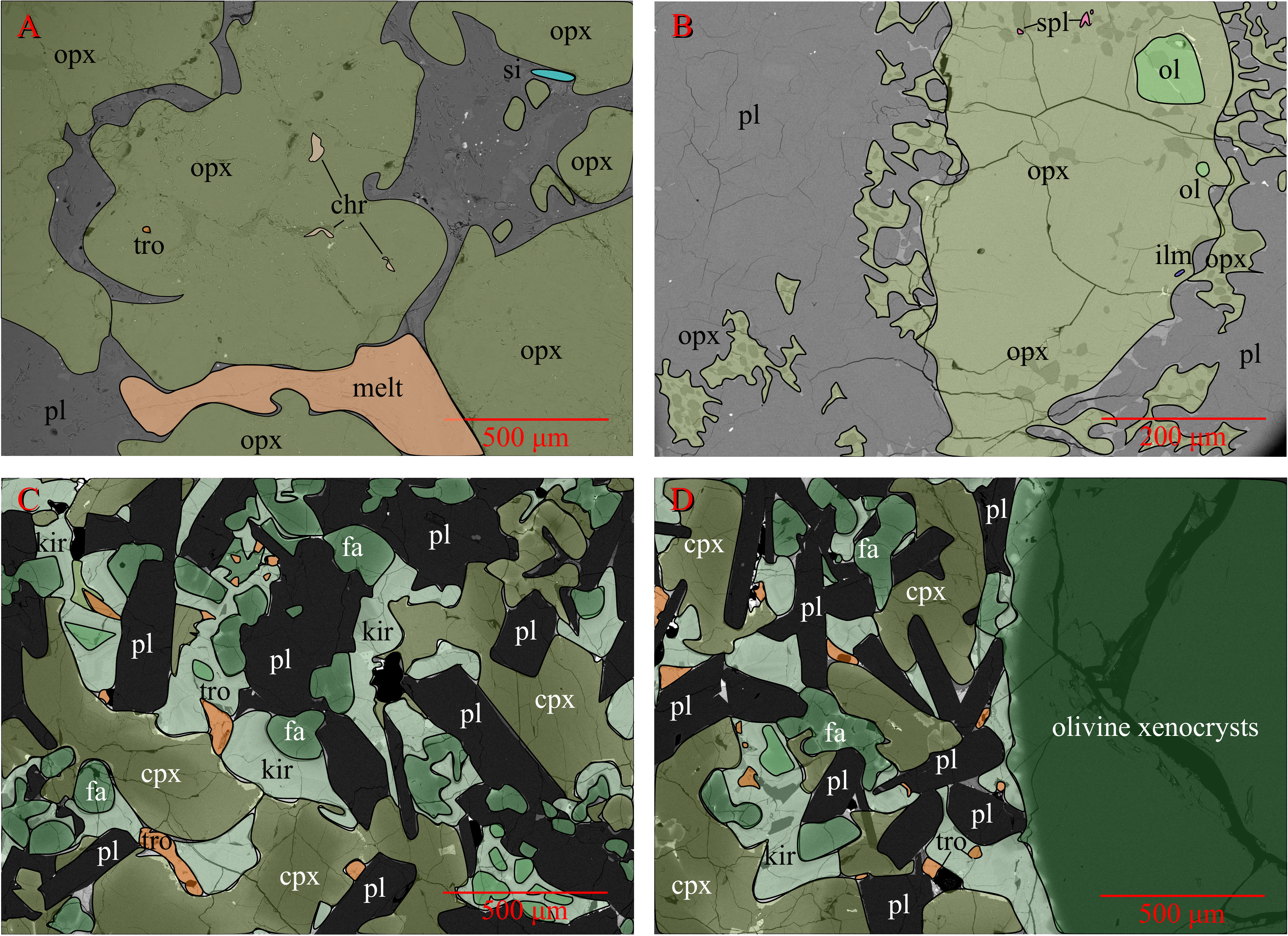}
\caption{\textbf{Backscatter electron (BSE) imaging of meteorite Arguin  $002$(A), Al Bir Lahlou (ABL) $001$(B),Oued Namous (ON) $001$(C and D).} Minerals are labeled as: pl, plagioclase; opx, orthopyroxene; tr, troilite; chr, chromite; si, silica; ol, olivine; ilm, ilmenite; spl, spinel; an, anorthite; cpx, clinopyroxene; kir, kirschsteinite; fa, fayalite. Color coding is a guide to the eye.}
\label{fig:meteorite BSE}
\end{figure*} 
\begin{figure*}[t]
\includegraphics[width=0.95\textwidth]{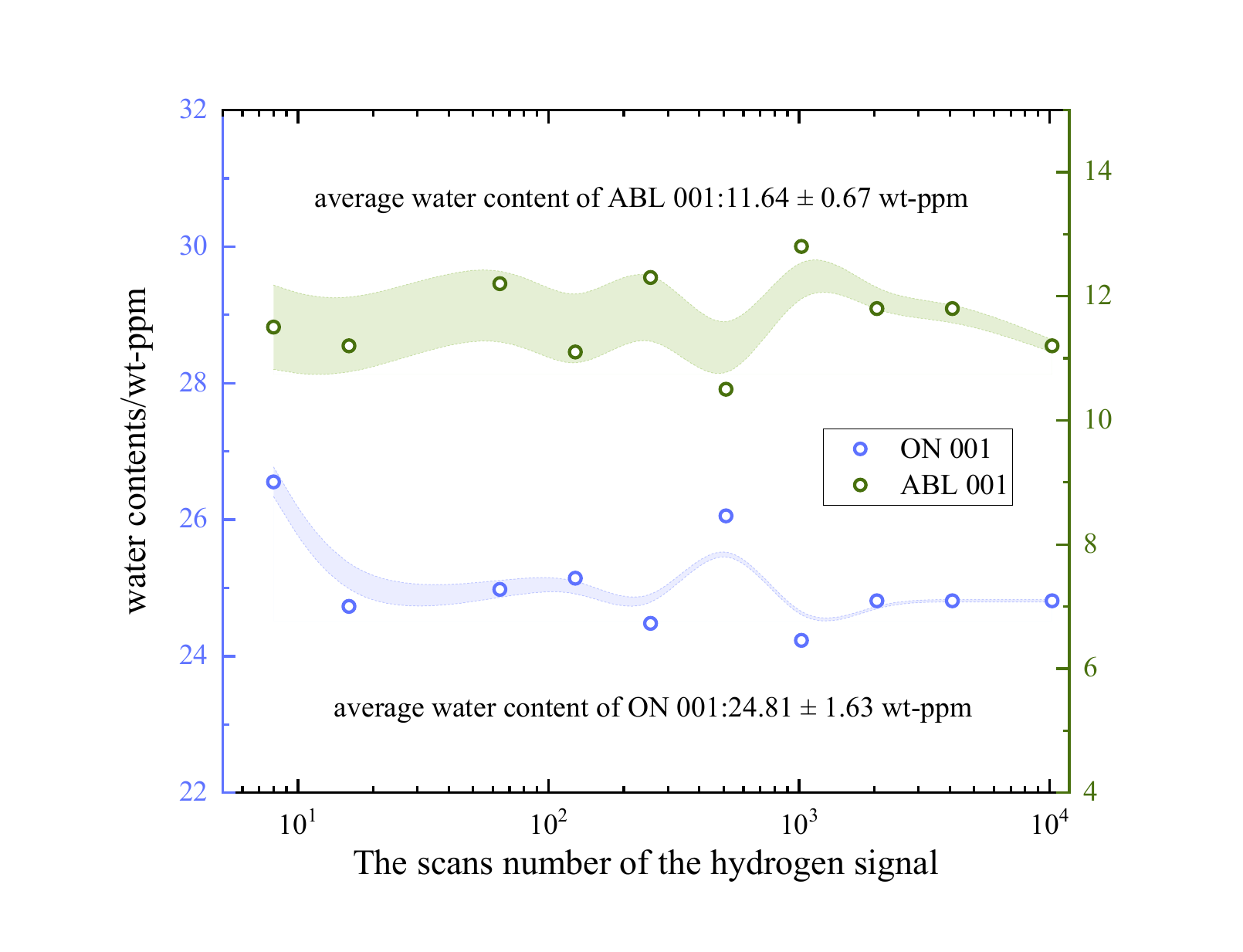}
\caption{\textbf{ the relationship between the water content of anorthites from meteorite ON 001 and ABL 001 and the amount of accumulated scans. }. The purple dots and green dots represent data from ON 001 and ABL 001, respectively. The purple and green areas represent the margin of error.}  
\label{fig:ABL001 and ON 001}
\end{figure*}
\begin{figure}[h]
\includegraphics[width=1\columnwidth]{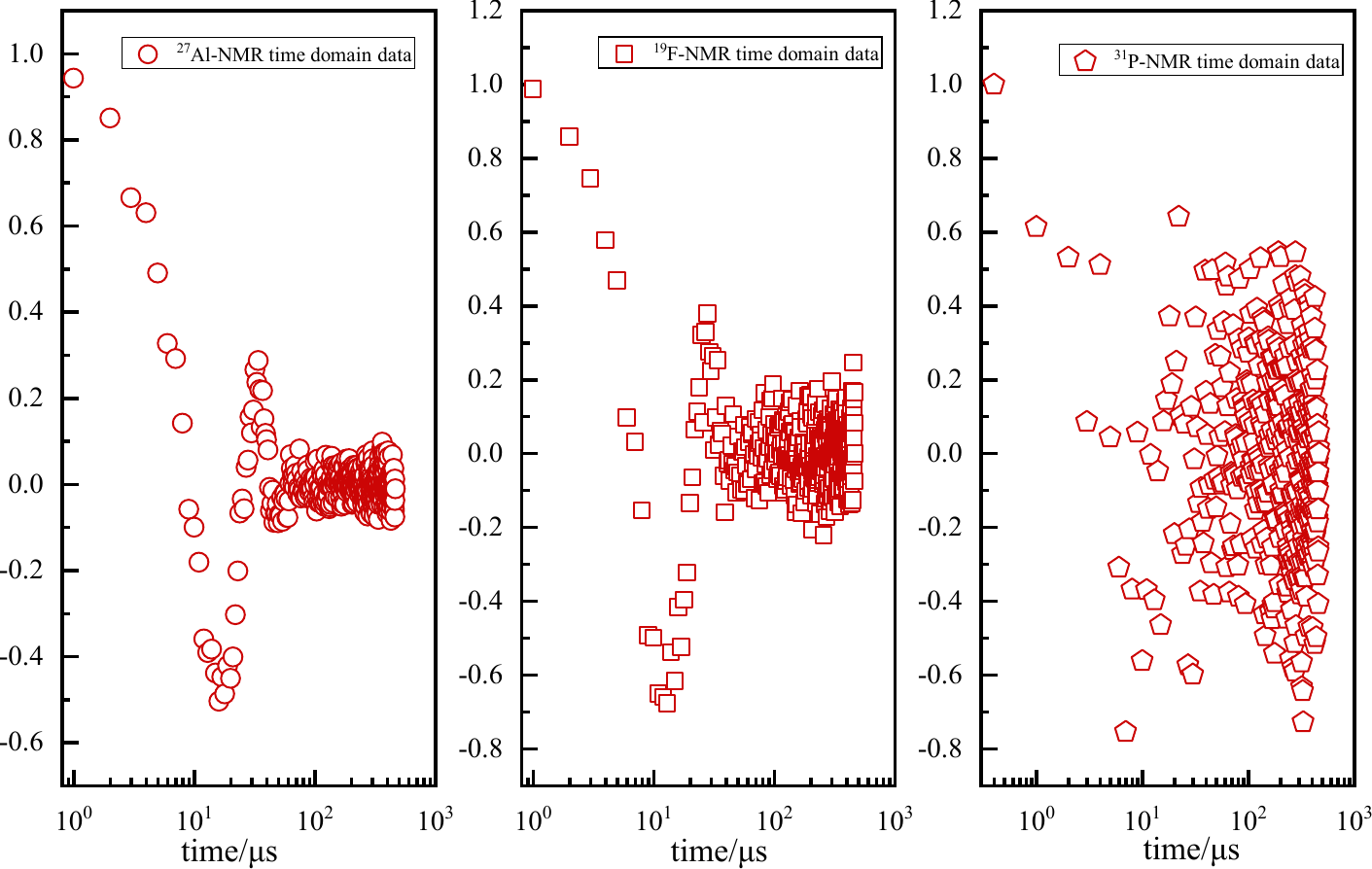}
\caption{\textbf{$^{27}Al$-NMR,$^{19}F$-NMR and $^{31}P$-NMR time-domain data acquired from anorthites of meteorite ON 001 under ambient conditions }. }
\label{fig: ON 001 trace element.pdf  }
\end{figure}
\begin{figure}[h]
\includegraphics[width=0.95\columnwidth]{Supplementary material/FTIR.pdf}
\caption{\textbf{Polarized IR spectra of anorthites from meteorite ABL 001 and ON 001, normalized to $1~mm$ sample thickness.}}
\label{FTIR}
\end{figure}
\begin{figure}[h]
\includegraphics[width=0.95\columnwidth]{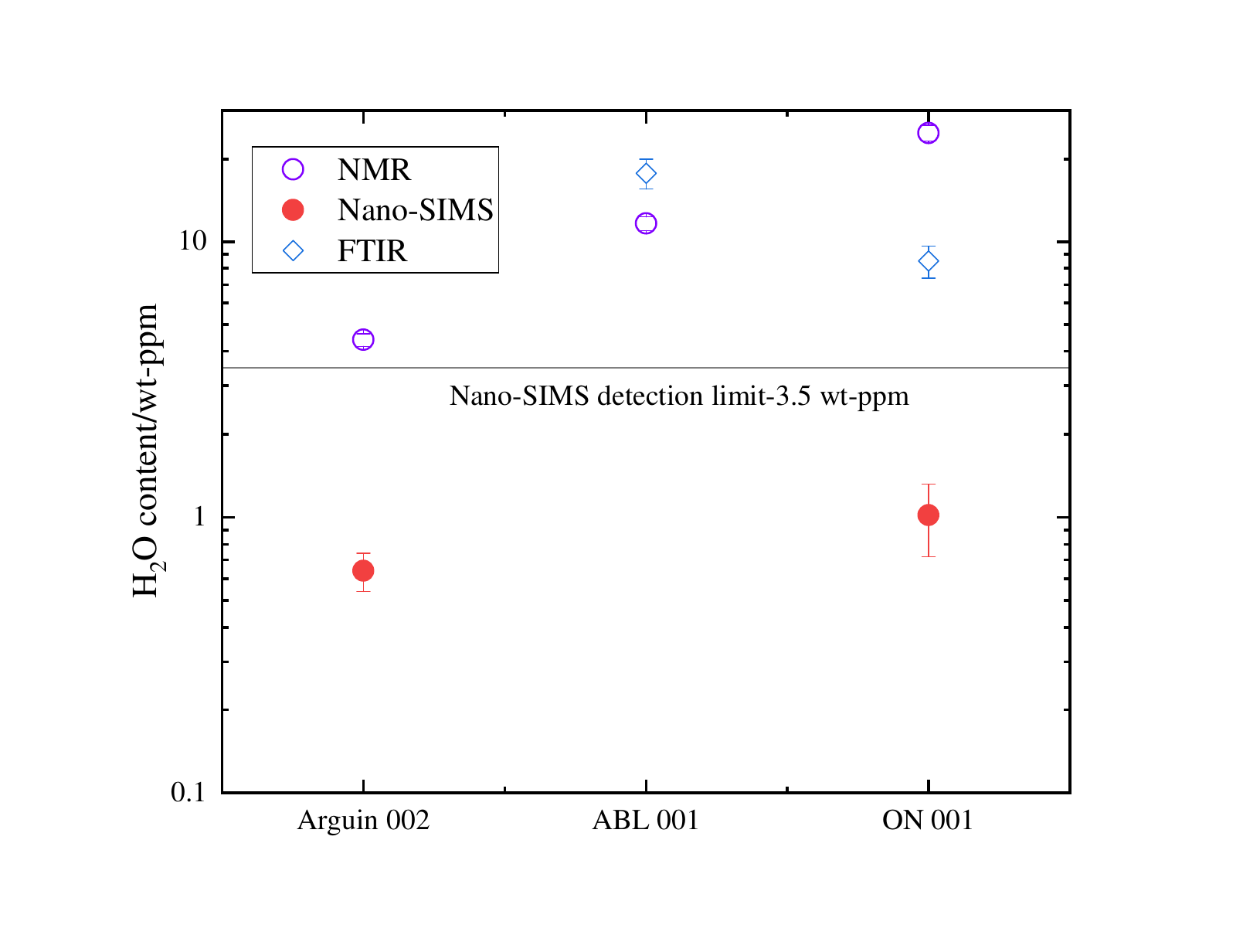}
\caption{\textbf{Water contents of meteorite Arguin 002, ABL 001 and ON 001 analyzed by NMR, FTIR and Nano-SIMS.}}
\label{fig.3 meteorite NMR FTIR SIMS data  }
\end{figure}
\begin{figure}[h]
\includegraphics[width=0.8\columnwidth]{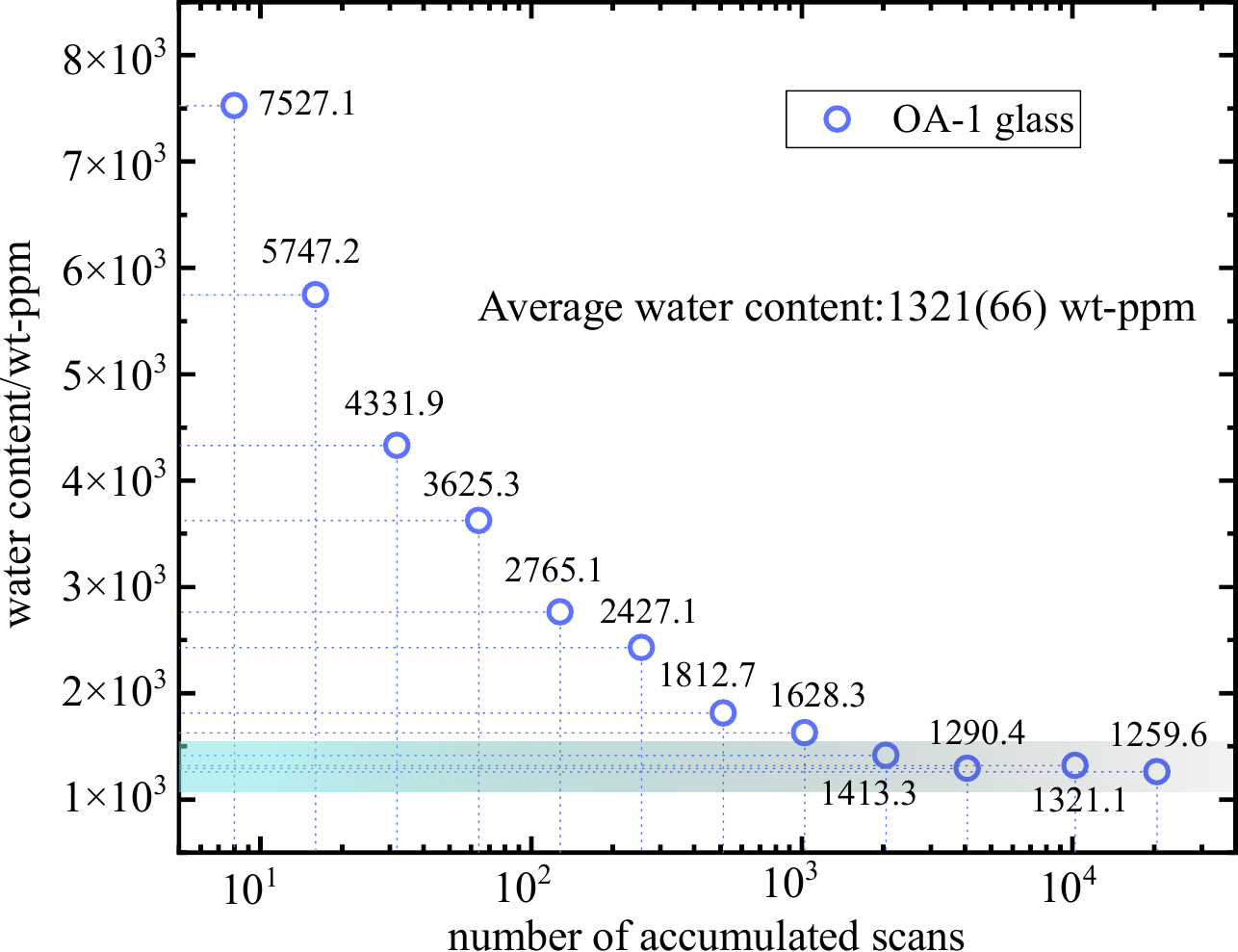}
\caption{\textbf{the relationship between the water content of OA-1 glass and the amount of accumulated scans. }}
\label{fig.OA glass  }
\end{figure}
 \FloatBarrier
 \section{Detailed NMR Measurements}
\subsection{rhyolitic glass OA-1}  
Accounting for rare and minor rhyolitic glass, an excitation coil instead of a commercial rotor as the sample chamber was used to avoid crushing glass and transmitting signals. The small solenoid coil(about $2~mm$ diameter,$4$ turns)  was made from $ 500~\mu m$ copper wire. Then the coil was connected to the home-built NMR probe equipped with adjustable capacitors for tuning frequencies to the desired resonances and impedance matching\cite{Fukushima2018}.\\
Pulse nutation experiments were performed and optimized $\pi/2$ pulse lengths were found to be $4~\mu s~at~27~W~and~1~\mu s~at~27~W$ for $^{27}Al$ and $^1H$.To allow sufficient relaxation of the spin systems, $T_1$ and $T_2$ experiments in part were carried out with long scan repetition times( spin-lattice relaxation time $T_1 \approx 180~ms$ and the spin-spin relaxation time  $T_2 \approx 130~\mu s$). Single Gauss pulse and boxcar pulse were conducted to acquire signals for $^{27}Al$ and $^1H$ nuclei. Their scan-normalized signal-to-noise ratios were found to be $\zeta_{Al}=4.48$ and  $\zeta_H=0.67 $ respectively at identical digitization frequencies. Given the estimated nuclear density ratio of about $n_H/n_{Al}=0.043$, the water concentration of glass was calculated to be  $1321(66)~wt-ppm$ (Supplementary material Figure \ref{fig.OA glass  }), which is in excellent agreement with the average data of FTIR and NanoSIMS.\\
\subsection{$LaH_3$ and $Al$  metal mixtures}  
We attempted to use single boxcar pulses and accumulated scanning times $1024$ to measure optimized SNR values for $^1H$ signals. Then $^{27}Al$ measurements were carried out in a series of accumulated scanning times from $8$ to $131073$ using single boxcar pulses. The signal-to-noise ratio of $^{27}Al$ increases with the square root of incremental accumulated scanning times. Relative hydrogen concentration is found to be as a function of the SNR ratio at fixed $^1H$ SNR and reliable values, within $\pm 1\sigma$ confidence levels, are approached after $\hat{\zeta}\approx 10^1$ has been achieved. It turns out that the aluminum normalized hydrogen content has an apparent linear relationship with optimal $^{27}Al$ to $^1H$ SNR ratio within the range from $0.1\%$ to $100\%$ (Supplementary material Figure\ref{fig.H calibrations LaH3 and Al  }). Our measurements suggest the detection limit of hydrogen content could be extrapolated to less than $10~wt-ppm$ or even $1~wt-ppm$ and indicate that NMR spectroscopy can be used to study NAMs with ultralow water content.\\
\subsection{NAMs-Stishovite}
NMR microresonator was used to encapsulate stishovite grains on account of the limited abundance. It was designed based on the magnetic flux tailoring Lenz lenses which have been used in in situ high-pressure NMR measurements in diamond anvil cells\cite{Meier2017, Meier2022}. Magnetic flux tailoring Lenz lenses were made from gold foils, which were cut into geometry shape ( $2.5~mm$ diameter and $25~\mu m$ thick) and drilled a hole ( $300~\mu m$ diameter) in the center and a $25~\mu m~ wide$ slit by laser cutting devices. The $300~\mu m$–by–$350~\mu m$ cylindrical sample chamber (diameter $\times$ thickness)  was composed of gold foils and an h-BN chip ( $2.5~mm$ diameter~and~$300~\mu m$ thick) with a hole ( $300~\mu m$  diameter). It was sealed by two additional h-BN chips. The resulting $20~nl$ cavity was filled with stishovite grains and the filling factor was almost close to $1$ \cite{Li2023}. It is worth noticing that stishovite is almost free of aluminum, thus we used the sample chamber volume instead of the sample volume to estimate the water content. Given the above-mentioned geometry of the resonator setup, as well as present spectrometer parameters ( $8192$ scans, $20~ MHz$ digitization frequency), the signal-to-noise value was found to be $132$. We constrained a total amount of about $3.32$ × $10$$^{15}$ hydrogen atoms present in the sample and determined the water concentration $6200(1500)~wt-ppm$.



\bibliography{references.bib}